\documentstyle[prb,preprint,aps,psfig]{revtex}
\tighten
\begin{document}
\draft
\title
{\bf  Cooper Instability in the Occupation Dependent Hopping Hamiltonians}
\author{H. Boyaci$^1$ and I. O. Kulik$^{1,2}$}
\address{
$^1$ Department of Physics,
Bilkent University,
Ankara 06533, Turkey \\
$^2$ B. Verkin Institute for Low Temperature 
Physics and Engineering, Nat. Acad. Sci. of Ukraine, Lenin Ave. 47, 310164
Kharkov, Ukraine}
\maketitle
\begin{abstract}
A generic Hamiltonian, which incorporates the effect of 
the orbital contraction  
on the hopping amplitude between the nearest
sites, is studied 
both analytically at the weak coupling limit and numerically at the
intermediate and strong coupling regimes for finite atomic cluster.
The effect 
of the orbital
contraction due to hole localization at atomic sites
is
specified with two coupling 
parameters $V$ and $W$ (multiplicative and additive contraction terms).
The singularity of the vertex part of the 
two-particle Green's function determines the 
critical temperature $T_c$ and the relaxation rate $\Gamma(T)$ of the order 
parameter at temperature above $T_c$. Unlike in conventional 
BCS superconductors, $\Gamma$ has a non-zero imaginary part which 
may influence the fluctuation conductivity of superconductor 
above $T_c$. We compute the 
ground state energy as a function of the particle number and magnetic flux 
through the cluster, and show the existence of the parity gap $\Delta$ 
appearing at the range of system parameters consistent with the appearance of 
Cooper instability. Numeric calculation of the Hubbard model
(with $U>0$) at arbitrary
occupation does not show any sign of superconductivity in small cluster.
\end{abstract}
\vspace{.5cm}

\narrowtext

\section{Formulation of the model}
High temperature superconductivity in the lanthanum \cite{bednorz1}, 
yttrium \cite{wu1} and related copper-oxide compounds remains a subject 
of intensive investigation and controversy. It was suggested that 
electron-phonon interaction mechanism, which is very successful in 
understanding of conventional (``low temperature") superconductors
within the Bardeen-Cooper-Schriffer scheme \cite{bardeen1}, may not
be adequate for high-$T_c$ cuprates, and even the conventional Fermi liquid
model of metallic state may require reconsideration. This opens an area
for investigation of mechanisms of electron-electron interaction which 
can be relevant in understanding peculiarities of superconducting, as well
as normal state, properties of cuprates. Specific to all of them is 
the existence of oxide orbitals. Band calculations 
\cite{hybertsen1,mahan1} suggest
that hopping between the oxygen $p_x,p_y$ orbitals and between the 
copper $d_{x^2-y^2}$ 
orbitals may be of comparable magnitude. On the experimental side, 
spectroscopic studies \cite{nucker1,kuiper1} clearly show that the oxygen band 
appears in the same region of oxygen concentration in which  
superconductivity in 
cuprates is the strongest. Therefore there exists a possibility
that specific features of oxide compounds may be related to oxygen-oxygen
hopping, or to the interaction between the copper and the rotational $p_x-p_y$ 
collective modes. If the oxygen hopping is significant then it 
immediately follows that intrinsic oxygen carriers ($p_x,p_y$ oxygen holes)
should be different from the more familiar generic $s$-orbital derived itinerant
carriers. The difference is related to low atomic number of oxygen 
such that removing or adding of one electron to atom induces a
substantial change in the Coulomb field near the remaining ion and 
therefore results in the change 
of the effective radius of atomic orbitals near the ion.
This will strongly influence the hopping amplitude 
between  this atom and the atoms in its neighborhood. Such ``orbital 
contraction'' effect  represents a source of strong interaction which 
does not simply reduce to the Coulomb (or phonon) repulsion (or 
attraction) between the charge carriers. It was suggested 
by Hirsch and coauthors \cite{hirsch1,hirsch2,hirsch3}, and by the
present authors \cite{kulik1,kulik2,boyaci1,kulik3} that the occupation 
dependent hopping can have relevance to the appearance of superconductivity
in high-temperature oxide compounds. In the present paper, we investigate
the generic occupation-dependent hopping Hamiltonians with respect 
to peculiarities of the normal 
state, and to the range of existence of the superconducting state.
Theoretical investigation of Cooper instability is supplemented by 
numeric study of pairing and diamagnetic currents in finite atomic 
clusters. We study the effect of Cooper pairing between the carriers and 
show that at certain values and magnitudes 
of the appropriate coupling parameters,
the system {\em is} actually superconducting. The 
properties of such superconducting 
state are in fact only slightly different from the properties
of conventional (low-$T_c$) superconductors. 
Among those we so far can only mention 
the change in the fluctuation conductivity above or near the 
critical temperature $T_c$. Relaxation of the pairing parameter to equilibrium 
acquires a small real part due to the asymmetry of contraction-derived
interaction between the quasi-particles above and below the Fermi energy.

Oxygen atoms in the copper-oxygen layers of the cuprates (Figure 1) 
have simple
quadratic lattice. We assume that $p_z$ orbitals of oxygen ($z$ is the
direction perpendicular to the cuprate plane) are bound to the 
near cuprate layers
whereas carriers at the $p_x,p_y$ 
orbitals may hop between the oxygen ions in the 
plane.

Let $t_1$ be the hopping amplitude of $p_x \, (p_y)$ and $t_2$ the
hopping amplitude of $p_y \, (p_x)$ oxygen orbitals between the nearest 
lattice sites  in the $x \, (y)$ direction in a square lattice 
with a lattice parameter $a$. Then the non-interacting Hamiltonian is
\begin{equation}
H_0=-t_1 \sum_{<ij>_x} a_i^{\dagger}a_j - t_2 \sum_{<ij>_y} a_i^{\dagger}a_j
-t_1 \sum_{<ij>_y} b_i^{\dagger} b_j - t_2 \sum_{<ij>_x}b_i^{\dagger} b_j
\end{equation}
where $a_i^{\dagger} \, (a_i)$ is the creation (annihilation) operator 
for $p_x$ and correspondingly $b_i^{\dagger} \, (b_i)$ for $p_y$ orbitals.
The interaction Hamiltonian includes the terms
\begin{equation}
H_1=\sum_{<ij>} a_i^{\dagger}a_j \left[ V m_i m_j + W (m_i+m_j) \right]
+ \sum_{<ij>} b_i^{\dagger}b_j \left[V n_i n_j + W (n_i +n_j) \right]
\end{equation}
where $n_i=a_i^{\dagger}a_i$, $m_i=b_i^{\dagger}b_i$. This corresponds
to the dependence of the hopping amplitude on the occupation numbers
$n_i, \, m_i$ of the form
\begin{equation}
\left( \hat{t}_{ij} \right)_{a_i \rightarrow a_j} = \tau _0 (1-m_i)(1-m_j)
+\tau _1 \left[ (1-m_i)m_j+m_i(1-m_j) \right] + \tau_2 m_i m_j
\end{equation}
and correspondingly $(\hat{t}_{ij})_{b_i \rightarrow b_j}$ of the same  form
with $m_i$ replaced with $n_i$. The amplitudes $\tau_0, \, \tau_1, \, \tau_2$
correspond to the transitions between the ionic configurations of oxygen:
\begin{eqnarray}
\tau_0: \, & O_i^-+O_j^{2-} & \rightarrow O_i^{2-}+O_j^- \nonumber\\
\tau_1: \, & O_i+O_j^{2-}   & \rightarrow O_i^{2-}+O_j \\
\tau_2: \, & O_i+O_j^-      & \rightarrow O_i^- +O_j \nonumber
\end{eqnarray}
$O$ corresponds to the neutral oxygen ion whereas $O^-$ to the single
charged and $O^{2-}$ to the double charged negative ions. Since oxygen atom has 
$1s^22p^42s^2$ configuration in its ground state,
filling of the $p$ shell to the full 
occupied configuration $2p^6$ is the 
most favorable. Amplitudes $V$ and $W$ relate 
to the parameter $\tau_0, \, \tau_1, \, \tau_2$ according to 
\begin{equation}
V=\tau_0-2\tau_1+\tau_2, \:
W=\tau_1-\tau_2 \, .
\end{equation}
Assuming $t_1=t_2$ and replacing $a_i, \, b_i$ with $a_i$
with the pseudo-spin indices 
$\sigma=\downarrow , \,  \uparrow$ we write the Hamiltonian Eq.(1) in the 
form
\begin{equation}
H=-t\sum_{<ij> \, {\sigma}} a_{i \, \sigma}^{\dagger}a_{j \, \sigma}
+H_U+H_V+H_W
\end{equation}
where
\begin{eqnarray}
H_U &=&U \sum_i n_{i \uparrow} n_{i \downarrow}  \\
H_V &=& V \sum_{<ij> \, \sigma} a_{i \sigma}^{\dagger} a_{j \sigma} n_{i, \bar{\sigma}}
n_{j, \bar{\sigma}} \\
H_W &=& W \sum_{<ij> \, \sigma} a_{i \sigma}^{\dagger} a_{j \sigma} (n_{i, \bar{\sigma}}
+n_{j, \bar{\sigma}})
\end{eqnarray}
where we also included the in-site Coulomb interaction (U) 
between the dissimilar
orbitals at the same site. $\sigma$ can also be considered as a real spin 
projection of electrons at the site. In that case, the pairing will originate
between the spin-up and spin-down orbitals, rather than between $p_x$ and $p_y$
orbitals. More complex mixed spin -and orbital- pairing 
configurations can also be
possible within the same idea of orbital contraction (or expansion) at
hole localization but are not considered in this paper. 
The following discussion 
does not distinguish between the real spin and the pseudo-spin pairing.
The Hamiltonian, Eq.(6), is a model one which can not refer to the reliable 
values of the parameters appropriate to the oxide materials. The purpose of our
study is rather to investigate the properties of superconducting transition
specific to the model chosen and to find the range of the
$U, \, V, \, W$ values which may correspond to superconductivity.
This will be done along the lines of the standard BCS model \cite{abrikosov1}
in the weak coupling limit, $U, \, V, \, W \, \rightarrow \, 0$, and by an
exact diagonalization of the Hamiltonian for a finite atomic cluster at 
large and intermediate coupling.

In the momentum representation, the Hamiltonian becomes
$H=H_0+H_1+H_2$ with
\begin{eqnarray}
H_0=\sum_{{\bf p} \, \sigma} \xi_{\bf p} a_{{\bf p} \sigma}^{\dagger} 
a_{{\bf p} \sigma}\\
H_1=\frac{1}{4} \sum_{p_1 p_2 p_3 p_4, 
\alpha  
\beta \, \gamma \, \delta}
a_{{\bf p}_1 \alpha}^{\dagger} a_{{\bf p}_2 \beta}^{\dagger} 
\Gamma_{\alpha \beta
\gamma \delta}^0(p_1, p_2 ,p_3 ,p_4) a_{{\bf p}_4 \delta} 
a_{{\bf p}_3 \gamma}
\end{eqnarray}
where 
\begin{equation}
\xi_{\bf p} = -t\sigma_{\bf p}-\mu,  \;\;\: \sigma_{\bf p}=2(\cos p_xa+
\cos  p_y a),
\end{equation}
and $\mu$ is the chemical potential. $\Gamma_{\alpha \beta \gamma \delta}^0$
is the zero order vertex part defined as
\begin{equation}
\Gamma_{\alpha \beta \gamma \delta}^0(p_1, p_2, p_3, 
p_4)= \left[ U+(W+
\frac{1}{2}\nu V) (\sigma_{{\bf p}_1}
+\sigma_{{\bf p}_2}+\sigma_{{\bf p}_3}+\sigma_{{\bf p}_4}) \right] 
\tau_{\alpha \beta}^x
\tau_{\gamma \delta}^x(\delta_{\alpha \gamma} \delta_{\beta \delta} - 
\delta_{\alpha \delta} \delta_{\beta \gamma}) \delta_{{\bf p}_1+{\bf p}_2, 
{\bf p}_3+{\bf p}_4}
\end{equation}
where $\tau_{\alpha \beta}^x$ is a Pauli matrix 
\begin{eqnarray*}
\left( \begin{array}{lr}
0 & 1 \\ 1 & 0 \end{array}  \right) .
\end{eqnarray*}
For reasons which will be clear later, 
we separated $H_V$  and put some part of it into 
the $H_1$ term, while the remaining part is included in the
$H_2$ term, thus giving
\begin{equation}
H_2=V \sum_{<ij>\sigma} a_{i\sigma}^{\dagger} a_{j\sigma}
(a_{i\bar{\sigma}}^{\dagger}
a_{i\bar{\sigma}}-\frac{\nu}{2})(a_{j\bar{\sigma}}^{\dagger}a_{j\bar{\sigma}}
-\frac{\nu}{2})
\end{equation}
with $\nu=<n_i>$ being the average occupation of the site.

\section{The Cooper instability in the occupation-dependent 
hopping Hamiltonians}
The Cooper instability realizes at certain temperature $T=T_c$ as a singularity
in a two-particle scattering amplitude at zero total momentum. Let's introduce 
a function
\begin{equation}
\Gamma(p_1 p_2,\tau - \tau^{ \prime})=<T_{\tau} a_{{\bf p}_1 \uparrow} 
(\tau) a_{-{\bf p}_1 
\downarrow}(\tau) \bar{a}_{-{\bf p}_2 \downarrow}(\tau^{ \prime}) 
\bar{a}_{{\bf p}_2 \uparrow}
(\tau^{ \prime})>
\end{equation}
where $\bar{a}_{{\bf p}\alpha}(\tau) = 
\exp(H\tau) a_{{\bf p} \alpha}^{\dagger} \exp(-H\tau)$, $a_{{\bf p}\alpha}
=\exp(H\tau) 
a_{{\bf p}\alpha} \exp(-H\tau)$ are the imaginary time $(\tau)$ creation and
annihilation operators. 
At ${\bf p}_1=-{\bf p}_2$, ${\bf p}_3=-{\bf p}_4$, 
the kernel of $\Gamma_{\alpha \beta \gamma \delta}$ is 
proportional to $G_{\alpha \beta}^x$ $G_{\gamma \delta}^y$ ($G$ is one-electron
Green function).
We keep notation  $\Gamma({\bf p p}^{\prime})$ for such a reduced Green function
specifying only momenta ${\bf p}={\bf p}_1=-{\bf p}_2$ and 
${\bf p}^{\prime}={\bf p}_3=-{\bf p}_4$. By assuming 
temporarily $V=0$, this Hamiltonian results in an equation for the Fourier 
transform $\Gamma(p,p^{\prime},\Omega)$
\begin{equation}
\Gamma({\bf p},{\bf p}^{\prime}, \Omega)= \Gamma^0({\bf p},
{\bf p}^{\prime})- T \sum_{\omega} \sum_{{\bf{k}}} \Gamma^0({\bf{p}},
{\bf{k}})G_{\omega}({\bf{k}})G_{-\omega+\Omega}(-{\bf{k}}) 
\Gamma({\bf{k}},{\bf{p}^{\prime}},\Omega)
\end{equation}
corresponding to summation of Feynmann graphs shown in Figure 2.
In the above formulas,
$\omega=(2n+1)\pi T$ and $\Omega=2\pi m T$
($n,m$ integers) are the discrete odd and even frequencies 
of the thermodynamic perturbation theory
\cite{abrikosov1}.
$G({\bf k}, \omega)$ is a one-particle Green function in a 
Fourier representation
\begin{equation}
G({\bf k}, \omega)= \frac{1}{\xi_k-i\omega}.
\end{equation}
Diagrams of Figure 2 are singular since equal momenta of two parallel
running lines bring together singularities of both Green functions 
$G({\bf k},\omega)$ and $G(-{\bf k}, \omega)$.

6-vertex interaction, Eq.(8), is not generally considered in the theories
of strongly-correlated fermionic systems. Such interaction also results in 
singular
diagrams for ${\bf p} \rightarrow -{\bf p}$ scattering shown in 
Figure 3. Since a closed loop in this figure does not carry any momentum 
to the vertex, it reduces to the average value of $\bar G$ which in turn is the 
average of the number operator, $<a^{\dagger} a>$. 
Taking into consideration of such diagrams is equivalent to 
replacing one of the $n_i$'s in Eq.(8) to its thermodynamical average 
$\nu=<a_{i \sigma}^{\dagger}a_{i \sigma}>$. Then the $V$ term 
can be added to the renormalized value of $W$, 
\begin{eqnarray*}
W \rightarrow W+ \frac{1}{2} \nu V
\end{eqnarray*}
We will check by numeric analysis in Sec. III 
to which extent such an approximation 
may be justified.

Solution to Eq.(16) can be received by putting 
\begin{equation}
\Gamma({\bf p},{\bf p}^{\prime}, \Omega)=A(\Omega)+B_1(\Omega)\sigma_{\bf p}
+B_2(\Omega)\sigma_{{\bf p}^{\prime}}+C(\Omega)\sigma_{\bf p} 
\sigma_{{\bf p}^{\prime}}.
\end{equation}
Substituting this expression into Eq.(16) and introducing the quantities
\begin{equation}
S_n(\Omega)=T\sum_{\omega}\sum_{\bf k} \sigma_{\bf k}^{n}G_{\omega}({\bf k})
G_{-\omega+\Omega}(-{\bf k})
\end{equation}
we receive a system of coupled equations for $A,B_1,B_2,C$
\begin{eqnarray}
\left( \begin{array}{cccc}
       1+U S_0 + \tilde{W} S_1 & US_1+{\tilde W}S_2 & 0 & 0 \\
        {\tilde W}S_0            & 1+{\tilde W} S_1 & 0 & 0 \\
        0 & 0 & 1+U S_0 +{\tilde W}S_1 & U S_1+{\tilde W} S_2 \\
        0 & 0 &  {\tilde W} S_0 & 1+{\tilde W} S_1 \end{array} \right) & &
\left( \begin{array}{c}
      A \\ B_1 \\ B_2 \\ C \end{array} \right)=  
\left( \begin{array}{c}
     U \\ \tilde{W} \\ \tilde{W} \\ 0 \end{array} \right) 
\end{eqnarray}
where $\tilde{W}=W+\frac{1}{2} \nu V$, which are solved to give
\begin{equation}
A=\frac{U-\tilde{W}^2S_2}{D} , \;
B_1=B_2=\frac{\tilde{W}(1+\tilde{W}S_1)}{D} , \;
C=-\frac{\tilde{W}^2S_0}{D} 
\end{equation}
where $D$ is a determinant
\begin{equation}
D=\left| \begin{array}{cc}
           1+US_0+\tilde{W}S_1 & US_1+\tilde{W}S_2 \\
           WS_0 & 1+WS_1 \end{array} \right| .
\end{equation}

The determinant becomes zero at some temperature which 
means an instability in the two-particle scattering 
amplitude ($\Gamma\rightarrow\infty$). This temperature is 
the superconducting transition temperature $T_c$. At $T_c$, Eq.(16)
is singular, which means that two-particle scattering amplitude gets infinite.
Below $T_c$, the finite value of $\Gamma$ is established by including the
non-zero thermal averages (the order parameters), 
$<a_{\bf p}^{\dagger} a_{-\bf p}^{\dagger}>$,
$<a_{\bf p} a_{-\bf p}>$. We first analyze the
case of non-retarded, non-contraction interaction $U$, and after that will
consider
the effect of the occupation-dependent hopping terms, $V$ and $W$.

\subsection{Direct non-retarded interaction}
Neglecting contraction parameters $V,W$, solution to Eq.(16) reduces to 
\begin{equation}
-\frac{1}{U}=T\sum_{\omega}\sum_{\bf k} \frac{1}{\xi_{\bf k}^2+\omega^2}
\end{equation}
which after the summation over the discrete frequencies reduces to the 
conventional 
BCS equation (at negative $U$)
\begin{equation}
\frac{1}{|U|}=\sum_{\bf k} \frac{1-2n_{\bf k}}{2\xi_{\bf k}},
\end{equation}
with $n_{\bf k}=(\exp(\beta\xi_{\bf k})+1)^{-1}$.
At finite frequency $\Omega$, Eq.(23) reduces to 
\begin{equation}
\ln \frac{T}{T_c}=T \sum_{\omega}\int_{-E_1}^{E_2} d\xi \frac{-i \Omega}
{(\xi^2+\omega^2)(\xi+i\omega+i\Omega)}
\end{equation}
where we replaced for simplicity an integration over the Brillouin, zone
$\int d^3k$, by the integration over the energy assuming that the density
of states near the Fermi energy $\mu$ is flat. 
$-E_1$ and $E_2$ are the lower and upper limits of 
integration equal to $-4 t-\mu$ and $4 t-\mu$, respectively.
Such an approximation is not very bad since most singular
contribution to integral comes from the point $\xi_p=0$
where the integrand is the largest.

Above $T_c$, Eq.(25) determines the frequency of the order 
parameter relaxation
\cite{abrahams1,gorkov1,kulik4}. 
There is a small change in this frequency compared to the
BCS model in which limits of the integration $(-E_1,E_2)$ are symmetric
with respect to the Fermi energy, and small in comparison to $\varepsilon_F$, 
therefore we briefly discuss it now. 

To receive a real-time relaxation 
frequency, Eq.(25) needs to be analytically continued to a real frequency
domain from the discrete imaginary frequencies $i\omega_n=(2n+1)\pi i  T$ 
\cite{abrikosov1}.
Using the identity
\begin{equation}
T \sum_{\omega}\frac{1}{(\omega+i \xi_1)(\omega+i \xi_2)...(\omega+i \xi_n)}
=(-i)^n\sum_{i=1}^{n} \prod_{i\neq j} \frac{n(\xi_i)}{\xi_i-\xi_j}
\end{equation}
where $n(\xi)$ is a Fermi function $n(\xi)=(\exp(\beta \xi)+1)^{-1}$ gives
\begin{equation}
\ln \frac{T}{T_c}= \frac{i \Omega}{2} \int_{-E_1}^{E_2} 
\frac{\tanh \frac{\xi}{2T}}{\xi(2\xi + i \Omega)} d\xi
\end{equation}
where 
\begin{equation}
T_c=\frac{2 \gamma}{\pi} \sqrt{E_1E_2}\exp \left(- \frac{1}{N(\varepsilon_F)|U|} \right)
, \: \ln \gamma=C=0.577.
\end{equation}
$C$ is Euler constant. 
Analytic continuation is now simple: we change  $\Omega$ to 
$i(\omega-i \delta)$, $\delta=+0$, to receive a function which will be analytic 
in the upper half plane of complex $\omega$,
$Im \omega>0$. The order parameter relaxation 
equation becomes 
\begin{equation}
\left( \ln \frac{T}{T_c}-\frac{\omega}{4}\int_{-E_1}^{E_2}\frac{\tanh \frac
{\xi}{2T}}{\xi(\xi-\frac{\omega}{2}+i\delta)}d\xi\right) \Delta=0.
\end{equation}
At $\omega \ll T_c$ and $T-T_c \ll T_c$, the real and imaginary parts of  
Eq.(29) are easily evaluated to give
\begin{equation}
\left( T-T_c-\frac{\pi i \omega}{8 T_c}+ \omega \frac{E_1-E_2}{4E_1E_2}
\right) \Delta=0.
\end{equation}
Thus, the order parameter relaxation equation at $T > T_c$ becomes
\begin{equation}
(1+i \lambda) \frac{\partial\Delta}{\partial t} + \Gamma \Delta= 0
\end{equation}
where 
\begin{equation}
\Gamma = \frac{8}{\pi}(T-T_c), \:\: \lambda=\frac{2(E_1-E_2)}{\pi E_1E_2} T_c.
\end{equation}
In comparison to the BCS theory in which $E_1=E_2=\omega_D$ 
($\omega_D$ is the Debye frequency) and therefore $\lambda=0$, 
we receive the relaxation which has 
a non-zero ``inductive'' component, $-i \lambda \Gamma$. Typically, 
$E_1 \sim E_2 \sim \varepsilon_F$ and therefore 
$|\lambda|$ is a small quantity. It increases however near the 
low ($\nu \ll 1$) or near the maximal ($\nu \simeq 2$) 
occupation where $E_1$ or $E_2$ become small. Such mode of relaxation is 
specific to a non-retarded (non-phonon) interaction which is not 
symmetric near $\varepsilon_F$ and spans over the 
large volume of the ${\bf k}$-space
rather than is restricted to a narrow energy $\omega_D \ll \varepsilon_F$ 
near the Fermi energy.

\subsection{Occupation-dependent hopping instability and relaxation}
Neglecting direct interaction, we put $U=0$ in Eq.(22) and 
receive
\begin{equation}
-\frac{1}{\tilde{W}}=S_1(\omega) \pm \sqrt{S_0(\omega)S_2(\omega)}
\end{equation}
where at finite frequency $\omega$
\begin{equation}
S_n(\omega)=N(\varepsilon_F)T \sum_{\omega} \int_{-E_1}^{E_2} 
\left(\frac{\xi+\mu}{-t}\right)^n \frac{\tanh \frac{\xi}{2T}}{2\xi-\omega
+i \delta} d\xi .
\end{equation}
Putting $\omega=0$ we receive from Eq.(33) a transition temperature $T_c$. 
The equation has a solution at $\tilde{W} < 0$, $\mu < 0$, or at
$\tilde{W}>0$, $\mu>0$ (we assume that $t>0$).
Plus or minus sign is chosen to receive the maximal value of 
$T_c$ (the second solution corresponding to smaller $T$, 
then, has to be disregarded since at $T<T_c$ the order parameter will be finite
and therefore Eqs.(20)-(22) do not apply). This gives an expression for 
$T_c$
\begin{equation}
T_c=\frac{2\gamma}{\pi} \sqrt{E_1E_2} \exp \left[ \frac{E_1-E_2}{2|\mu|t}
(t-|\mu|)+\frac{E_2^2-E_1^2}{8\mu^2}\right] 
\exp \left(-\frac{t}{2|\tilde{W}|N(\varepsilon_F)}\right)
\end{equation}
where $\mu<0$, $\tilde{W}<0$ (second exponent is dominating the first one 
in the weak coupling limit $\tilde{W} \rightarrow 0$). 
Real and imaginary parts of $S_n(\omega)$ are calculated at $\omega \ll T_c$
\begin{eqnarray}
ImS_n(\omega) \simeq -\frac{\pi\omega}{8T_c}\left( -\frac{\mu}{t} \right)^n
N(\varepsilon_F)\\
ReS_n(\omega)=\frac{\omega}{4}N(\varepsilon_F) \left(-\frac{\mu}{t} \right)^n \times
\left\{ \begin{array}{ll} 
       \frac{E_2-E_1}{E_1E_2}, & n=0\\
       \frac{E_2-E_1}{E_1E_2}+ \frac{2}{\mu} \ln \frac{\gamma \sqrt{E_1E_2}}{T_c},
                               & n=1\\
       \frac{E_2-E_1}{E_1E_2}+\frac{E_1-E_2}{\mu^2}+
                               \frac{2}{\mu} \ln \frac{\gamma \sqrt{E_1E_2}}{T_c},
                               & n=2 \end{array} \right.
\end{eqnarray}
Equation for $\lambda$ 
is received with a value larger than the previous one (Eq.(32))
\begin{equation}
\lambda \simeq \frac{T_c}{\mu} \left( 3 \ln \frac{2 \gamma 
\sqrt{E_1E_2}}{\pi T_c}+ \frac{2 \mu(E_2-E_1)}{E_1E_2}+\frac{E_1-E_2}{2\mu}
\right).
\end{equation}
Eigenvalue equation gives the ${\bf p}$-dependence 
of the two particle correlator
$\Gamma({\bf p},{\bf p}^{\prime})= \\ <a_{{\bf p} \uparrow}^{\dagger} 
a_{-{\bf p} \downarrow}^{\dagger} a_{-{\bf p}^{\prime} \downarrow}
a_{{\bf p}^{\prime} \uparrow} >$
near $T_c$
\begin{equation}
\Gamma({\bf p}, {\bf p}^{\prime})=C \left[ S_2-S_1(\sigma_{\bf p} + 
\sigma_{{\bf p}^{\prime}}) + S_0 \sigma_{\bf p} \sigma_{{\bf p}^{\prime}} 
\right].
\end{equation}
Since $C$ diverges at $T_c$, this determines that order parameter
becomes macroscopic at $T<T_c$. Then, the 
pair creation operator, $a_{\bf p}^{\dagger} a_{-\bf p}^{\dagger}$, will
almost be a number, i.e., we may decompose Eq.(39) into a product 
\begin{equation}
\Delta_{\bf p}^{\star} \Delta_{\bf p} = <a_{{\bf p} \uparrow}^{\dagger}
a_{-{\bf p} \downarrow}^{\dagger}><a_{-{\bf p}^{\prime} \downarrow}
a_{{\bf p}^{\prime} \uparrow}> 
\end{equation}
and, to be consistent with the ${\bf p}$, ${\bf p}^{\prime}$ dependences,
by putting $\xi_{\bf p}=\xi_{{\bf p}^{\prime}}$ we receive 
\begin{equation}
\Delta_{\bf p}=C_1 \left( \exp(i \theta/2) \sqrt{S_2(0)}+\exp(-i \theta /2)
\sqrt{S_0(0)} \right) \exp(i\varphi)
\end{equation}
where 
\begin{equation}
\cos \theta= - S_1(0)/\sqrt{S_0(0)S_2(0)}
\end{equation}
and $\varphi$ is an overall phase which is irrelevant for a single 
superconductor but is important for calculating currents in multiple or 
weakly coupled superconductors. Therefore, system undergoes a pairing transition
at temperature found from the Eq.(35). Since the pairs are charged, the
state below $T_c$ can not be non-superconducting.

We have not calculated the Meissner response but in the following 
section we present
numerical calculation of flux quantization which supports the above statement.

\section{Exact diagonalization of the occupation-dependent hopping
Hamiltonians in finite cluster}
We calculate the ground state energy of a cubic system as shown in
Figure 4. A magnetic flux $\Phi$ is produced by a solenoid passing through the 
cube. Corners of the cube are the lattice sites, which can be  
occupied by electrons. With the inclusion of the magnetic flux, model
Hamiltonian, Eq. 6, becomes
\begin{eqnarray}
H=-t \sum_{<ij>\sigma} a_{i \sigma}^{\dagger} a_{j \sigma} \exp(i \alpha_{ij})+h.c.
+U \sum_i n_{i \uparrow}n_{i \downarrow} + \nonumber \\
+ \sum_{<ij>\sigma} 
a_{i \sigma}^{\dagger} a_{j \sigma} \left[ V n_{i \bar{\sigma}} 
n_{j \bar{\sigma}} + W (n_{i \bar{\sigma}}+n_{j \bar{\sigma}}) \right]
\exp(i \alpha_{ij}) +h.c.
\end{eqnarray}
where 
\begin{equation}
\alpha_{ij}=(2\pi/\Phi_0)\int_{{\bf r}_i}^{{\bf r}_j} {\bf A} \, d{\bf l}
\end{equation}
and $\Phi_0=hc/e$
is the magnetic flux quantum. Throughout the calculations we take $t=1$. 

We start with constructing the model Hamiltonian. In a Hilbert 
space of one electron 
\begin{eqnarray}
a=\left( \begin{array}{lr}
        0 & 1 \\ 0 & 0 \end{array}  \right), \; 
a^{\dagger}=\left( \begin{array}{lr}
        0 & 0 \\ 1 & 0 \end{array} \right) .
\end{eqnarray}
with a basis specified as $\psi_0=(0,1)$ for the ground state ($n=0$)
and $\psi_1=(1,0)$ for the excited state ($n=1$). In case of $N$ states
operator of annihilation $a_n$ takes the form
\begin{equation}
a_n=s^{n-1}\otimes a \otimes e^{N-n}
\end{equation}
where $e$ is the unit matrix and $s$ is unitary matrix
\begin{eqnarray}
e=\left( \begin{array}{lr}
         1 & 0 \\ 0 & 1 \end{array} \right), \;
s=\left( \begin{array}{lr}
         1 & 0 \\ 0 & -1 \end{array} \right) 
\end{eqnarray}
and $\otimes$ stands for the Kronecker matrix multiplication.
Explicitly, we have
\begin{eqnarray*}
a_1&=&a \otimes e \otimes e \otimes e \ldots \otimes e \\
a_2&=&s \otimes a \otimes e \otimes e \ldots \otimes e \\
 &\ldots& \\
a_N&=&s \otimes s \otimes s \ldots \otimes s \otimes a
\end{eqnarray*}
Thus, for example, for two states
\begin{eqnarray}
a_1=\left( \begin{array}{cccc}
          0 & 1 & 0 & 0 \\ 0 & 0 & 0 & 0 \\
          0 & 0 & 0 & 1 \\ 0 & 0 & 0 & 0 \end{array} \right), \;
a_2=\left( \begin{array}{cccc}
          0 & 0 & 1 & 0 \\ 0 & 0 & 0 &-1 \\
          0 & 0 & 0 & 0 \\ 0 & 0 & 0 & 0 \end{array} \right) .
\end{eqnarray}
These matrices, which are annihilation operators, and corresponding
Hermitian conjugate matrices, which are the creation 
operators, satisfy the Fermi anti-commutation relation.
These operators are sparse matrices with only $N/2$ non-zero 
elements, which are equal to $\pm 1$. Next we solve the Schr\"{o}dinger 
equation $H\psi=E\psi$. We implemented a novel algorithm 
for solving such sparse systems, which will be described elsewhere.

The cubic cluster within the Hubbard Hamiltonian and no external flux applied 
to the system was studied previously by Callaway et. al. \cite{callaway1}. 
Quantum Monte Carlo methods applicable to large systems 
within the Hubbard model (both attractive and repulsive), but not
the occupation-dependent hopping Hamiltonians, are reviewed in a paper of 
Dagotto \cite{dagotto1}. 

\subsection{The number parity effect}
Superconductivity reveals itself in the lowering of the ground state energy
as electrons get paired. Therefore the energy needs to 
be minimal for even number 
of electrons $n$ and will attain a larger value when $n$ is odd. We consider
a ``gap" parameter \cite{matveev1}
\begin{equation}
\Delta_l=E_{2l+1}-\frac{1}{2}\left(E_{2l}+E_{2l+2}\right)
\end{equation}
as a possible ``signature" of superconductivity
(where $E_m$ corresponds to the ground state energy for $m$ fermions). For all
interaction parameters set to zero ($U=V=W=0$), 
no sign of pairing is observed. To
check our analytic results of Sec. IIb and the argument following Eq.(34), 
we calculated $\Delta$ above and below the half-filling 
($n=8$ in case of cubic cluster). Below the half-filling
chemical potential is negative ($\mu<0$) and
above the half-filling it is positive ($\mu>0$). 
We first checked that the $W\rightarrow 0^+$,
$W\rightarrow 0^-$ and $V\rightarrow 0^+$, $V\rightarrow 0^-$ calculation
is consistent with an exact solution available for a non-interacting 
system of $n$ electrons. 

We then test our program for the 
case of negative-$U$ Hubbard Hamiltonian ($U<0, \, V=0,\, W=0$) which is 
known to be superconducting (e.g. Refs. 22,23). Positive-$U$ 
Hubbard model does not show any sign of superconductivity, in
disagreement with some statements in the literature \cite{pao1}. Our 
calculations 
can not disprove the (possible) non-pairing mechanisms of superconductivity
but these seem to be unlikely models for the problem of superconductivity
in oxides which clearly shows pairing of electrons (holes) in the
Josephson effect and in the Abrikosov vortices. The relation
$2eV=\hbar \omega$ is justified in the first case \cite{witt1} and 
flux quantum of a vortex is $hc/2e$ in the second \cite{gammel1}, both 
with the value of the charge equal to twice the electronic 
charge, $e$. 

Figure 5 shows the dependence of the ground state energy upon the
number of particles in case of negative-$U$ and positive-$U$ Hubbard models 
assuming $V=0$ and $W=0$. Such dependences are typical for any value of 
$|U|$. There clearly is the pairing effect when $U<0$ and there is no 
sign of pairing at $U>0$. 

Tests for pairing in the contraction
$V$, $W$-models ($V \neq 0, \, U=W=0$ and $W\neq 0,\,U=V=0$, respectively) 
are shown in Figs. 6,7.
The results are in agreement with our perturbative calculation of Sec. II 
and with its extension for the intermediate and strong coupling limits
$|V|\gtrsim t$, $|W|\gtrsim t$. Since chemical potential is negative below the 
half-filling and positive above the half-filling,
there is no pairing in the former case ($\tilde{W} \rightarrow 0^+$)
and there is a sign of pairing in the latter case ($\tilde{W} > 0$),
in accord with the value of the effective coupling constant $\tilde{W} =
W+\frac{1}{2}\nu V$.
Similarly, for $\tilde{W} \rightarrow 0^-$ below the half-filling there is a 
sign of pairing ($\Delta \neq 0$) while above the half-filling there 
is no pairing. These results are summarized in Table 1.

For larger values of the interaction
parameters, the perturbative results do not remain
applicable anymore. Figure 8b shows the dependence of the parity gap
$\Delta$ on the strength of the interaction. From Figure 8 it is
understood that the $W$ interaction introduces a ``signature" of pairing 
in a similar way as the negative-$U$ interaction does. The possibility 
of the ``contraction" pairing has been investigated formerly in the papers
\cite{hirsch3,boyaci1}.

\subsection{Flux quantization}
Flux quantization is another signature of superconductivity 
which is a consequence of the Meissner effect. We also tested for 
the periodicity of the energy versus 
flux dependence with the period $\Phi_1=hc/2e$ as compared to the period 
$\Phi_0=hc/e$ in the non-interacting system \cite{kulik6,buttiker1}.
Unfortunately, the even harmonics of $\Phi_0$-periodic dependence of the 
ground state energy (and related to it, the harmonics 
of the persistent current $J=-\partial E/
\partial \Phi$ \cite{kulik6,buttiker1}) may simulate the
pairing in a non-superconductive system.
Small-size (mesoscopic) system can mask the superconducting 
behavior \cite{dagotto1}. Flux quantization in Hubbard Hamiltonians was 
studied formerly in Refs. 29-31.

We first demonstrate the behavior of 
the ground state energy with respect to flux,
Figure 9. 
A 
characteristic feature
of mesoscopic system suggests that addition of one extra particle to the
system changes the sign of the derivative of the 
ground state energy with respect to
magnetic flux at $\Phi=0$. That is, depending on the parity of the number
of particles and on the number of sites, system can change from paramagnetic
to diamagnetic state or vice versa. But this behavior is not always observed
for the cubic geometry studied. 
Except the sign change from $n=2$ to $n=3$ and
from $n=7$ to $n=8$, no such behavior is seen. 
As mentioned above, however, the $\Phi_1$-periodic component
of the $E(\Phi)$ dependence begins to appear at the higher value
of $n$ (Figure 9,c). 
For both contraction parameters equal to zero, i.e. $W=V=0$, we observe 
appearance of the $hc/2e$-periodic component for some values of $U$
(Figure 10). Even for positive (repulsive) values of $U$, it is possible to 
see a local minimum appearing at $\Phi=hc/2e$ (Figure 10b). This is
in agreement with the authors' previous works \cite{boyaci1,ferretti1}. But 
this minimum, which does not lead to an exact periodicity of the 
ground state energy with a period 
$\Phi_0/2$, should not be attributed to superconductivity, this is
rather a characteristic behavior in mesoscopic systems.

For $U<0$ (while $W=V=0$), the expected mesoscopic behavior, 
that is the change of the sign of the slope of ground state energy 
at $\Phi=0$, starts to demonstrate itself (Figure 11). But this
happens at sufficiently large absolute values of (negative) $U$.
For other values of $U$,  however,  there is no such change.

More pronounced
$hc/2e$-periodic components are observed 
with the introduction of non-zero interaction parameters. 
The role of $W$ on 
the ground state energy, when both $U$ and $V$ are zero, is 
shown in Figure 12. Meanwhile setting both $U$ and $W$ to zero
and observing the effect of the non-zero $V$ shows that $V$ does not
play a role as significant as the other two interaction parameters do.
There is not much difference in the
behavior of the ground state energy upon magnetic flux  between the
zero and non-zero $V$ (for example $V=-1$) cases.

\section{Conclusions}
We studied the peculiarity of electron conduction 
in systems in which conduction band is derived 
from the atomic shells with a small number of electrons ($N_e$) in 
an atom. Such materials may include oxygen ($N_e=8$) in the oxides,
carbon ($N_e=6$) in borocarbides (e.g. $LuNi_2B_2C$), hydrogen
($N_e=1$) in some metals (e.g., $Pd-H$). Some materials of this 
kind are superconductors. It was argued that the Coulomb effects
within the atoms strongly influence the inter-atom wave function
overlap between the atomic sites and therefore the electron hopping amplitude
between the sites. The phenomenology of such conduction mechanism results in a
novel, to the conventional solid state theory, Hamiltonians
called the occupation-dependent-hopping (or contraction)
Hamiltonians, specified with the two coupling parameters $V$, $W$. We then 
attempted a study of superconductivity in such systems within 
the BCS-type approach assuming the Cooper pairing
of electrons. The weak-coupling limit allows determination of the range
of parameters $V$, $W$ values and also of the in-site Coulomb
interaction $U$ value which show the Cooper instability.
The strong-coupling limit was addressed by  a
numeric calculation on finite clusters using 
the novel algorithm (of non-Lanczos
type) for eigenvalues of large sparse matrices. One of the 
results of this numeric calculation was that the positive-$U$ Hubbard model,
sometimes believed to be a candidate for high-$T_c$
superconductivity, does not comply with the goal.

This work was partially supported by the Scientific and Technical 
Research Council of Turkey (T\"{U}B{\.I}TAK) through 
the BDP program.

\begin{figure}
\psfig{file=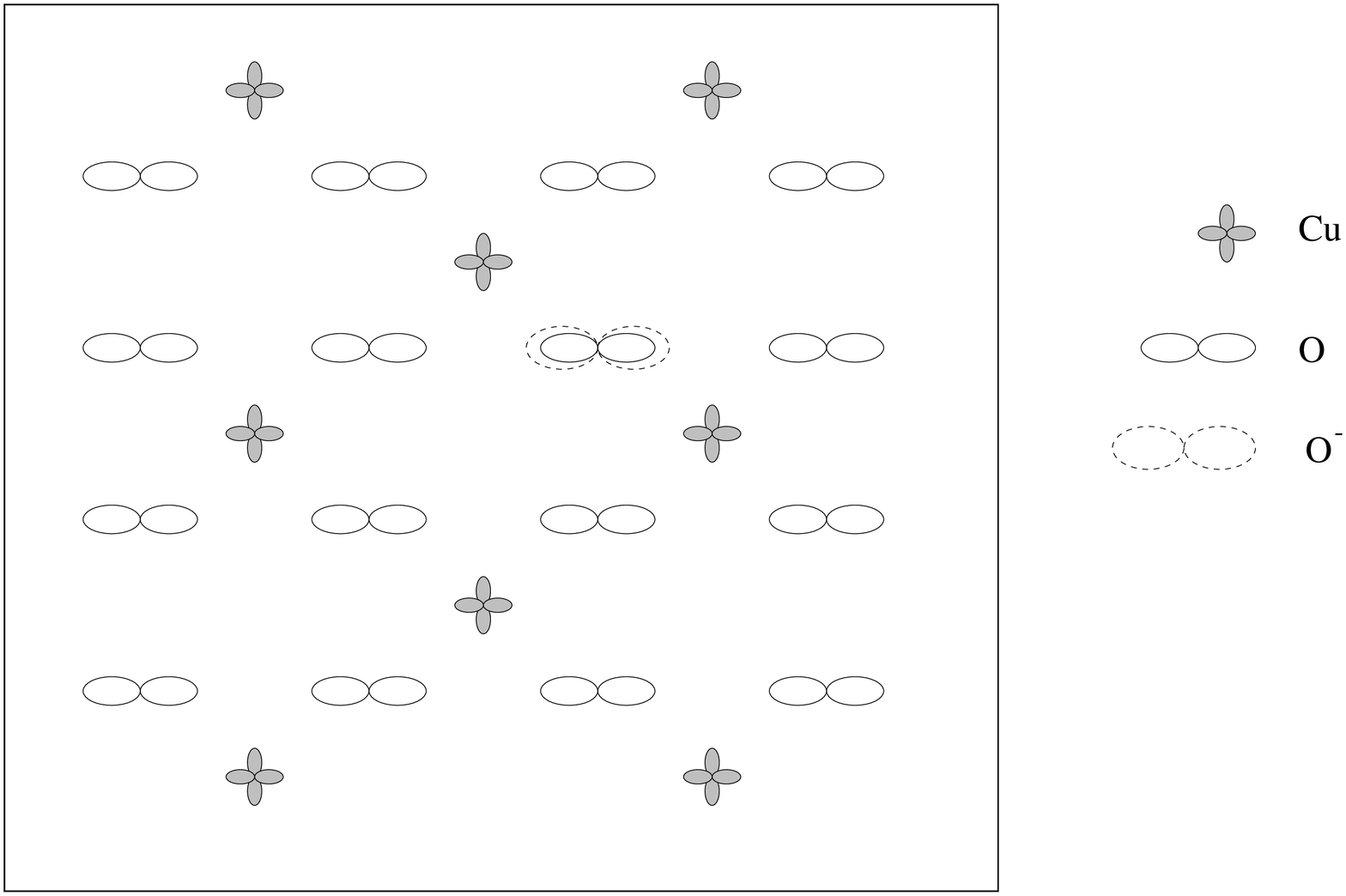,height=7.5cm,width=12cm}

\vspace{1cm}

\caption{Site configuration in the $CuO_2$ plane of cuprates. Dotted line
represents the effect of orbital contraction/extension due
to the localization/delocalization of an extra hole at 
a specific site. The enlarged  orbital attains the larger
value of the hopping amplitude to the nearest sites.}
\newpage
\end{figure}
\begin{figure}
\psfig{file=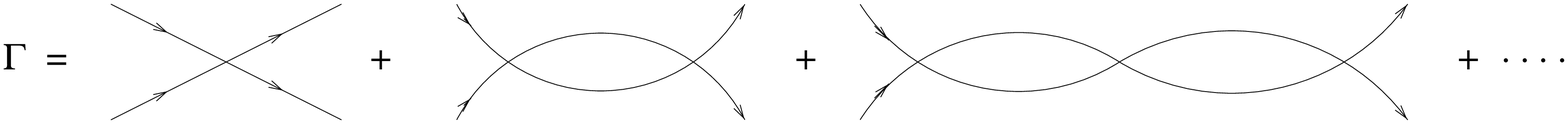,height=2cm,width=16cm}
\caption{Cooper diagrams for 4-vertex interactions, $U$ and $W$.}
\end{figure}
\newpage
\begin{figure}
\hspace{3.5cm} \psfig{file=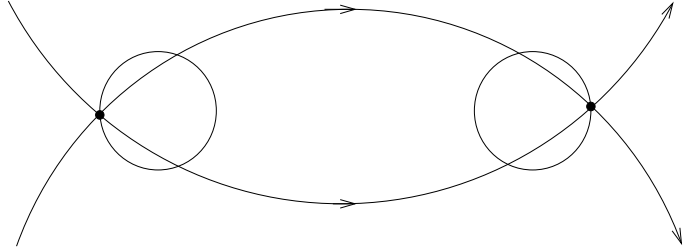,height=3.3cm,width=9cm}
\caption{Cooper diagram for 6-vertex interaction, $V$.}
\end{figure}
\newpage
\begin{figure}
\psfig{file=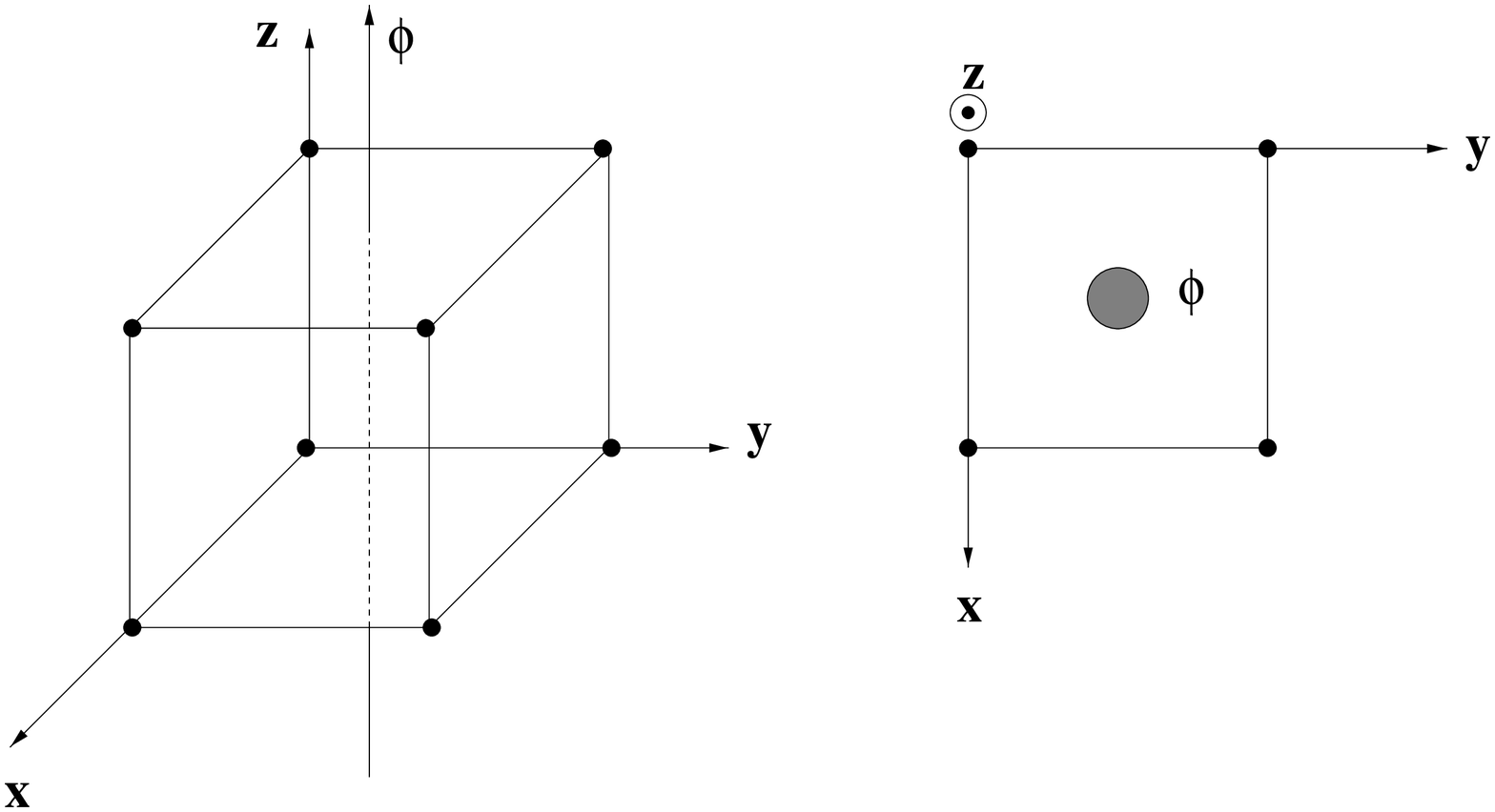,height=8cm,width=16cm}
\caption{Sample configuration. The flux through the cube is produced
by a solenoid.}
\end{figure}
\newpage
\begin{figure}
\psfig{file=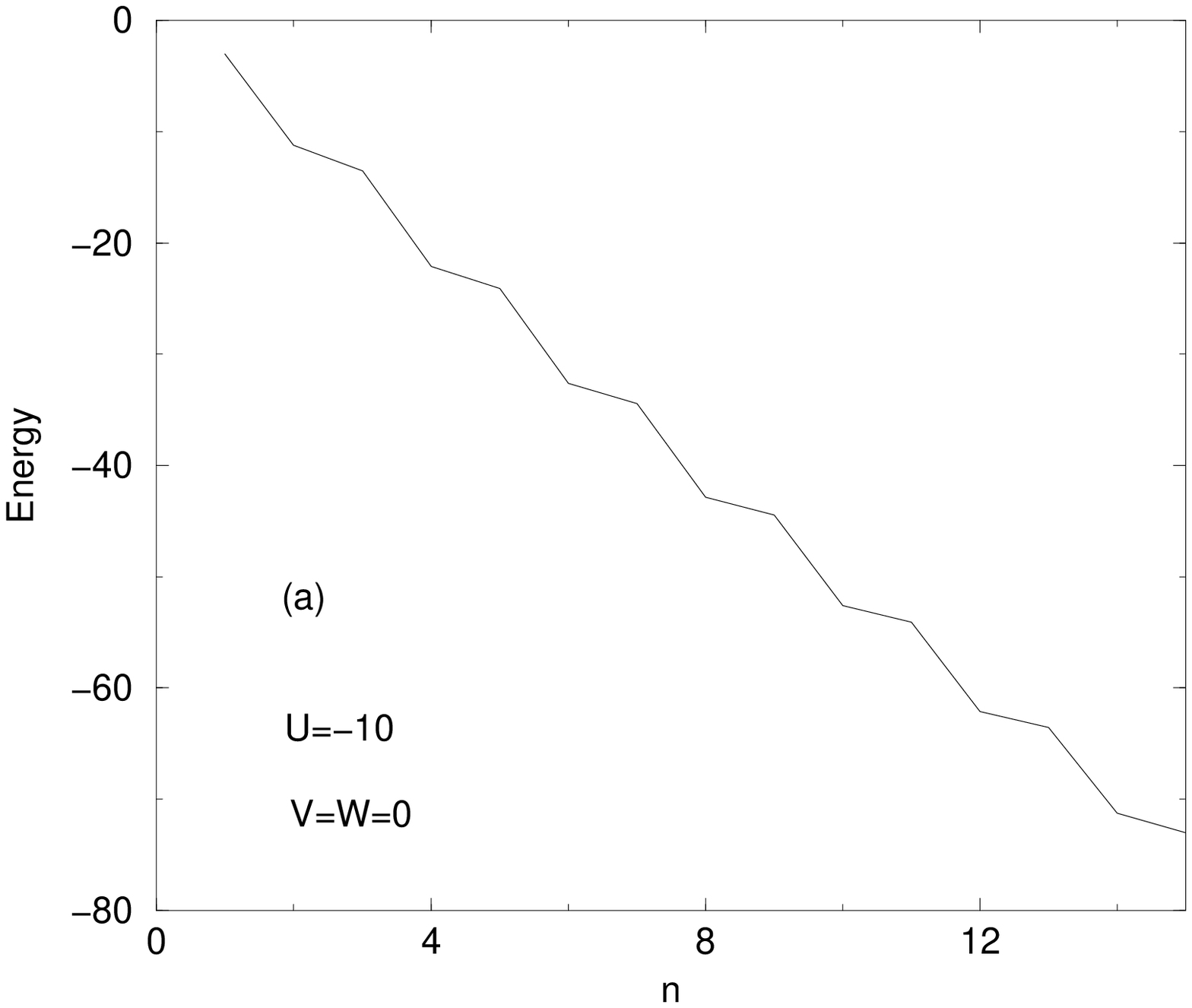,height=6cm,width=6cm}
\psfig{file=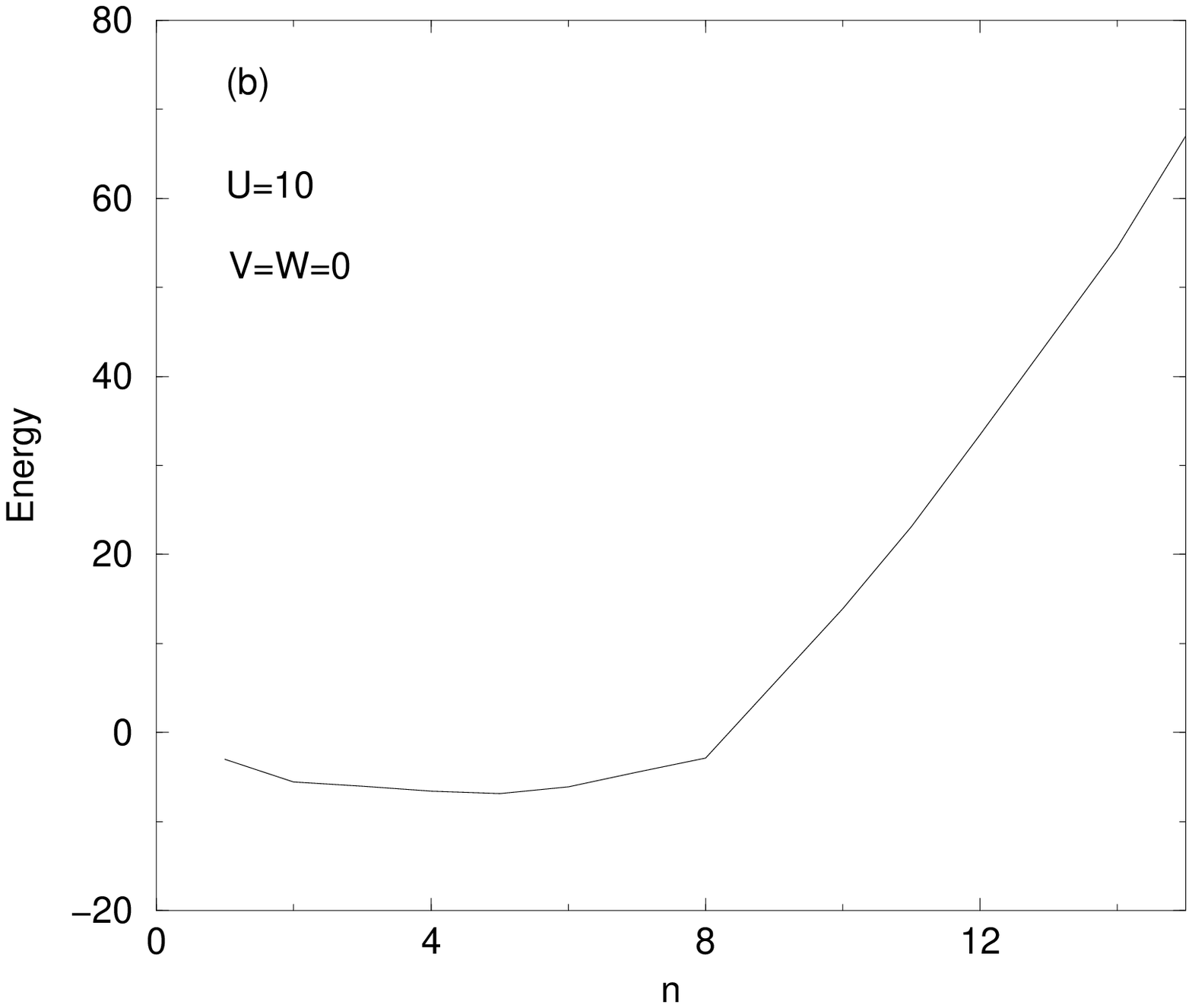,height=6cm,width=6cm}
\caption{Dependence of the ground state energy upon the number of particles
with $U\neq 0$ and $V=W=0$.
(a) For $U<0$, the pairing effect is clearly seen. (b) For $U>0$, there 
is no pairing}
\end{figure}
\newpage
\begin{figure}
\psfig{file=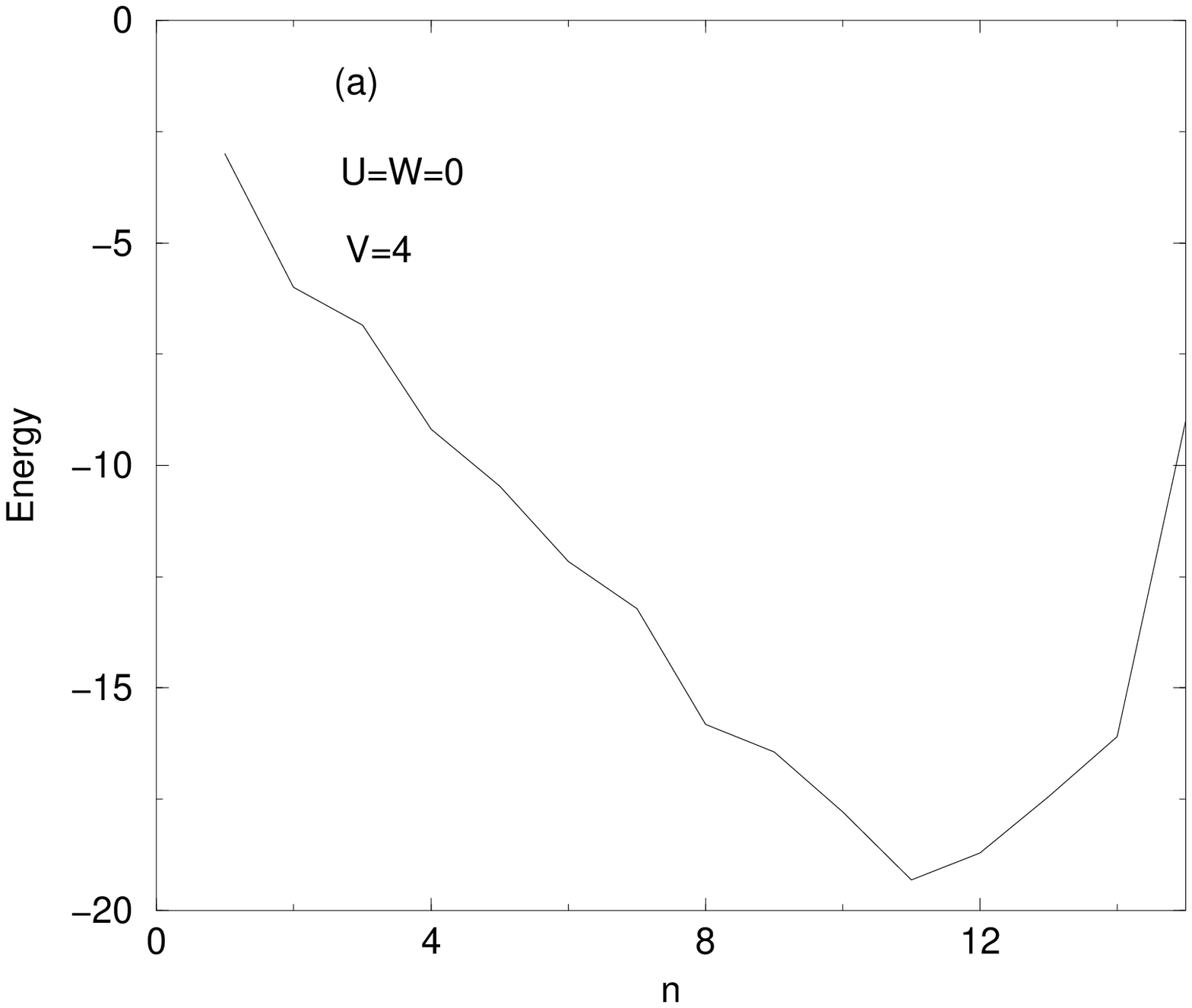,height=6cm,width=6cm}
\psfig{file=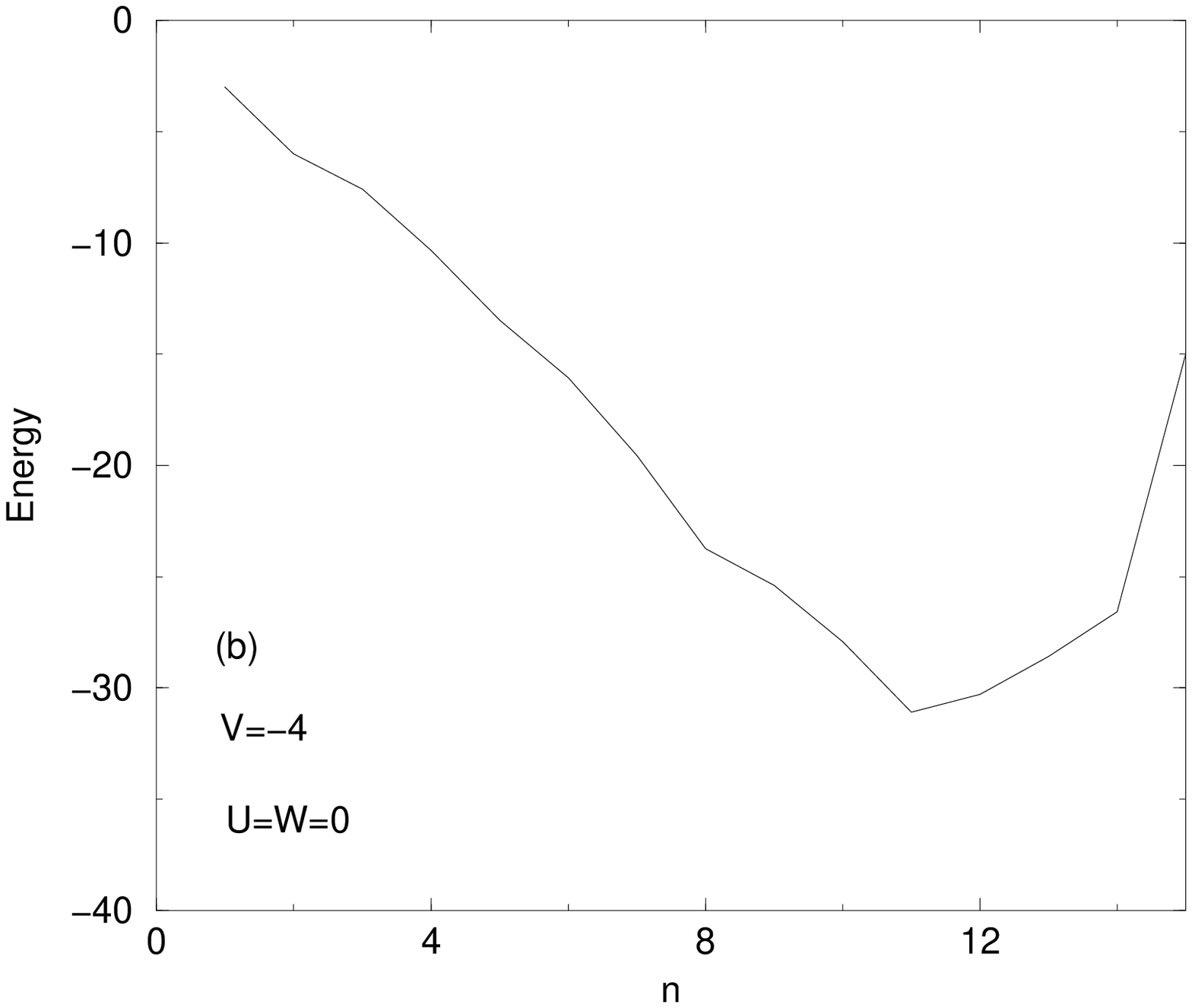,height=6cm,width=6cm}
\caption{Dependence of the ground state energy upon the number of particles 
with $V\neq 0$ and $U=W=0$.
(a), (b) Both for $V>0$ and $V<0$, around the half-filling, there is a small 
pairing effect.}
\end{figure}
\newpage
\begin{figure}
\psfig{file=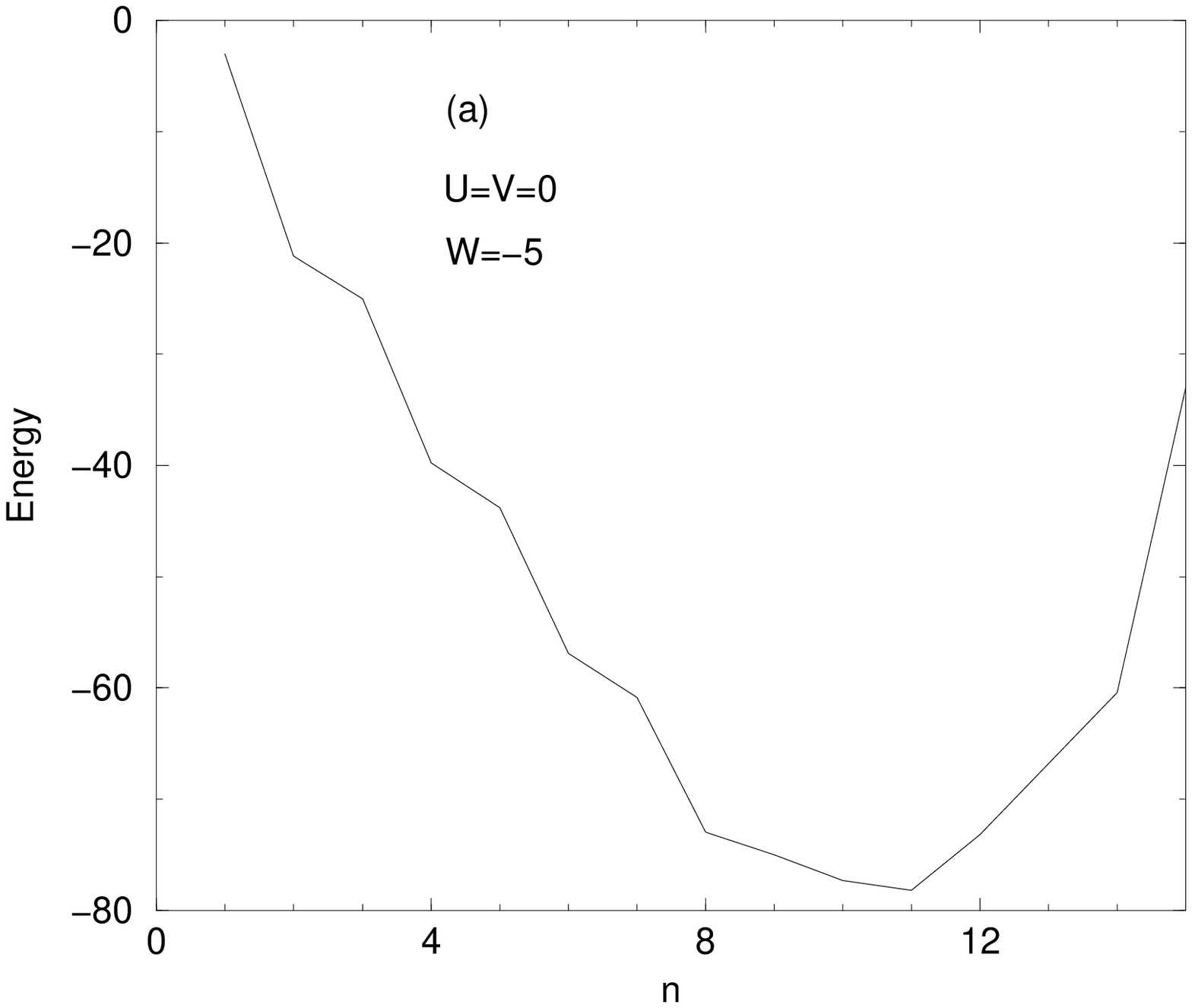,height=6cm,width=6cm} 
\psfig{file=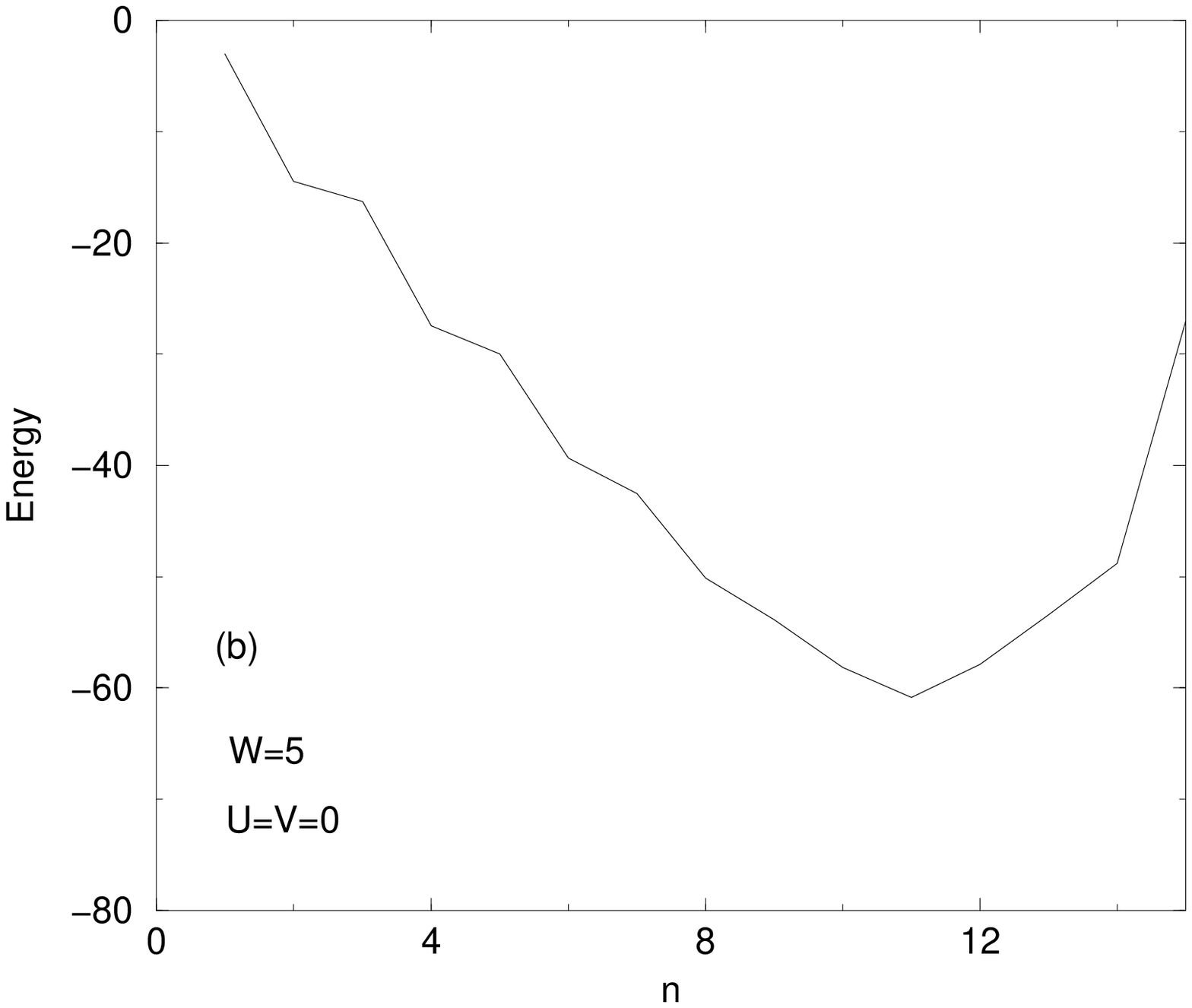,height=6cm,width=6cm} 
\caption{Dependence of the ground state energy upon the number of particles 
with $W \neq 0$ and $U=V=0$.
(a), (b) Both for $W>0$ and $W<0$, there is a more pronounced pairing 
effect below the half-filling.}
\end{figure}
\newpage
\begin{figure}
\psfig{file=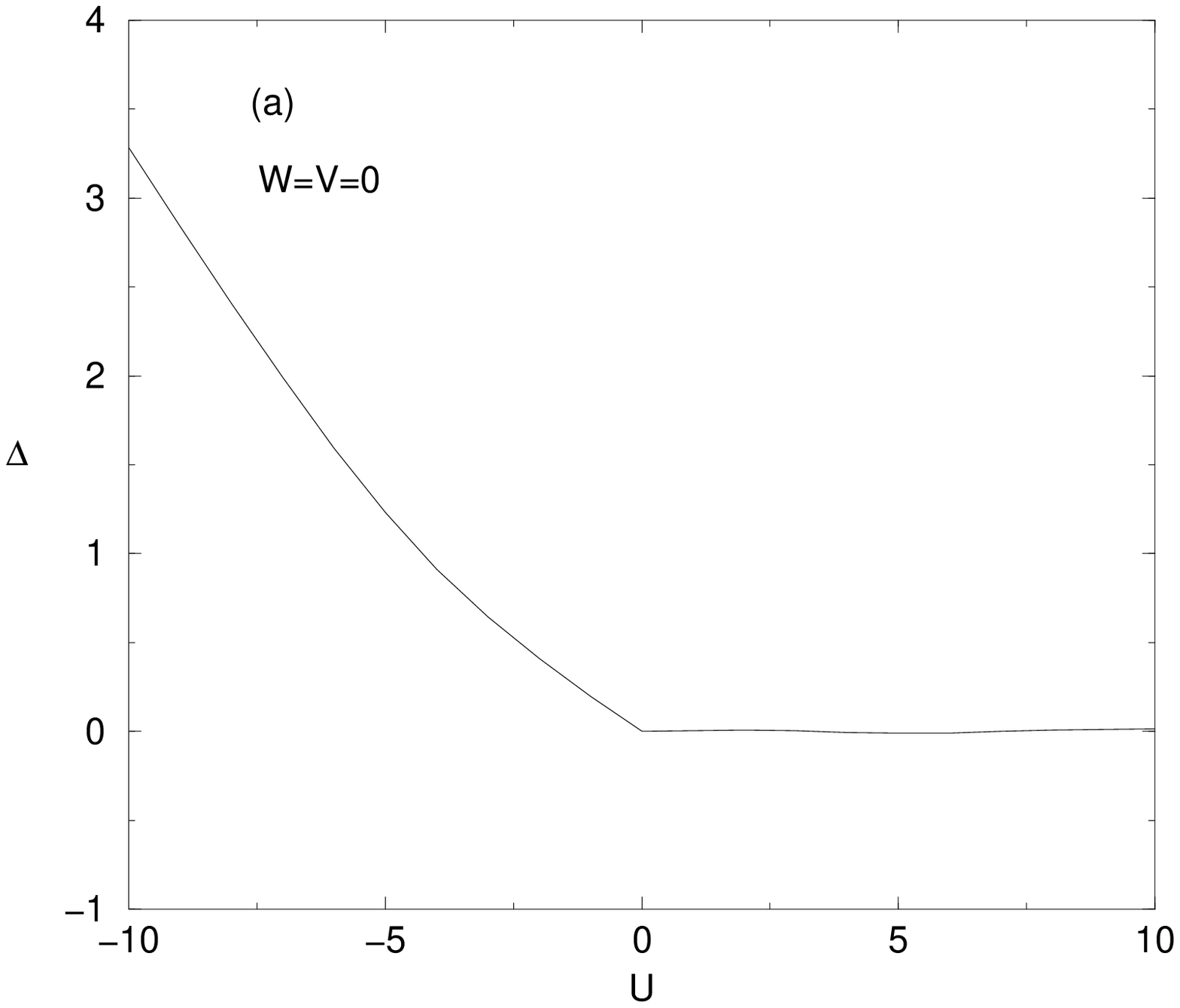,height=6cm,width=6cm}
\psfig{file=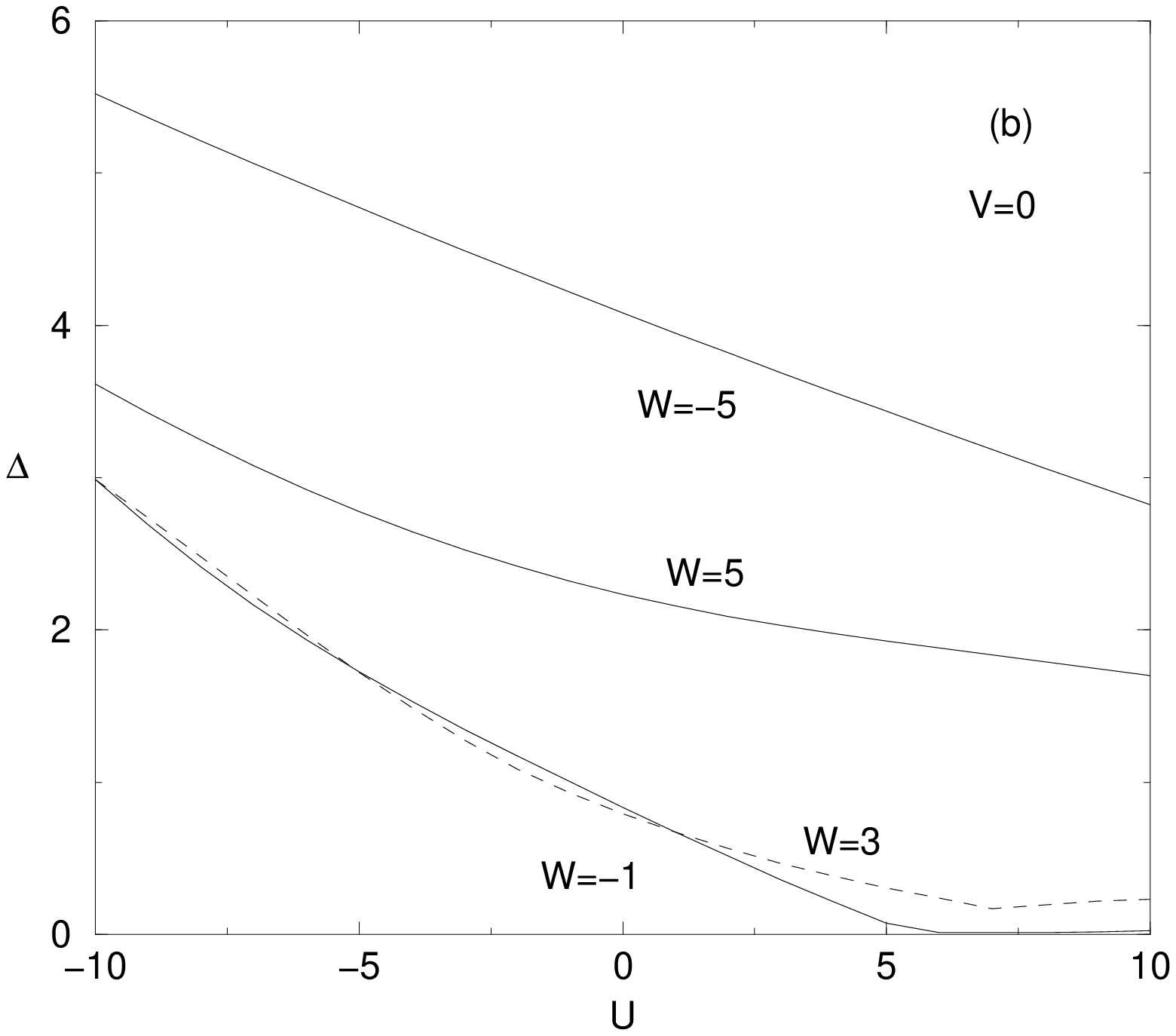,height=6cm,width=6cm}
\psfig{file=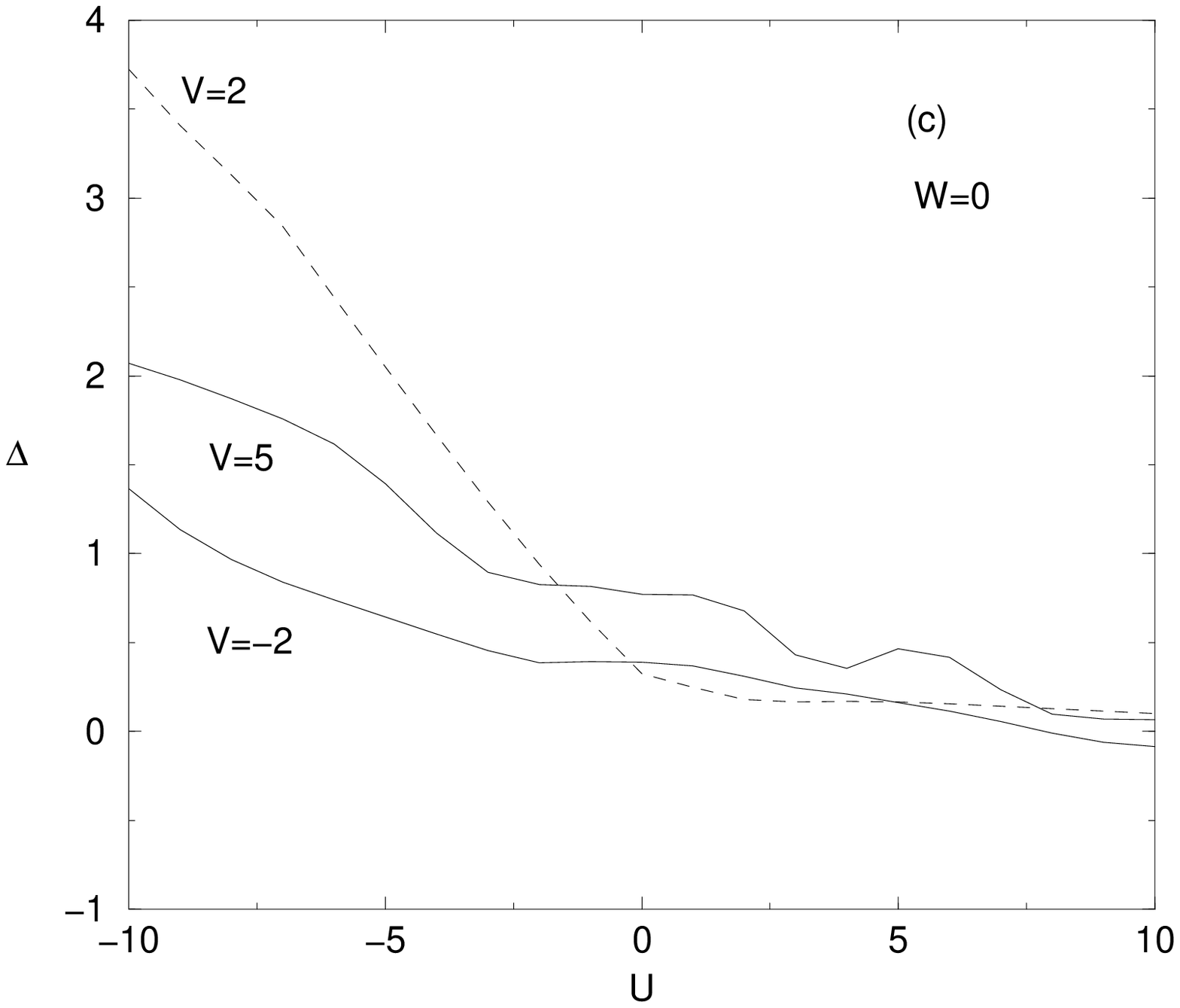,height=6cm,width=6cm}
\caption{Dependence of the parameter $\Delta$ upon $U$ for various values of
$W$ and $V$ below the half-filling.}
\end{figure}
\newpage
\begin{figure}
\psfig{file=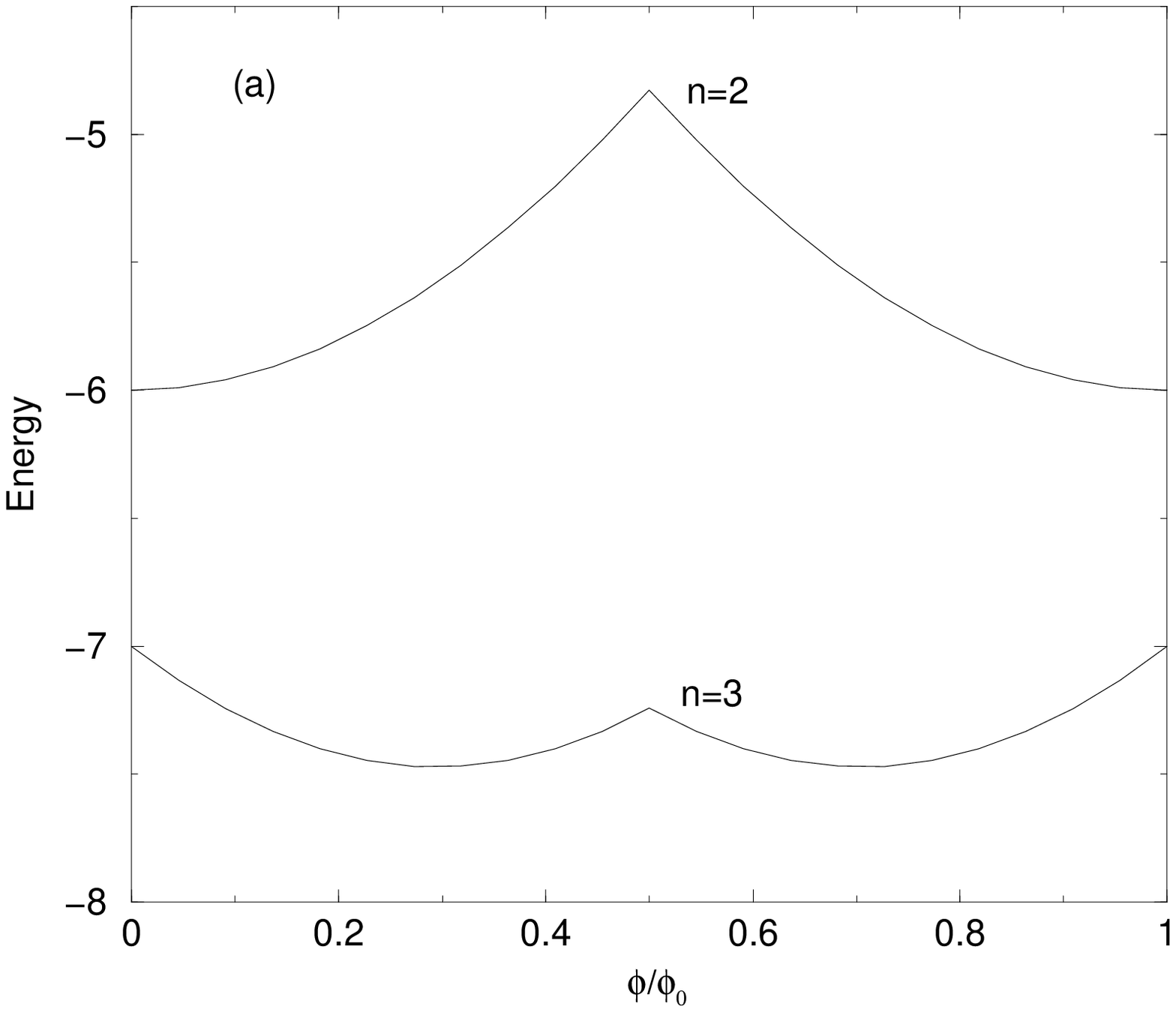,height=6cm,width=6cm}
\psfig{file=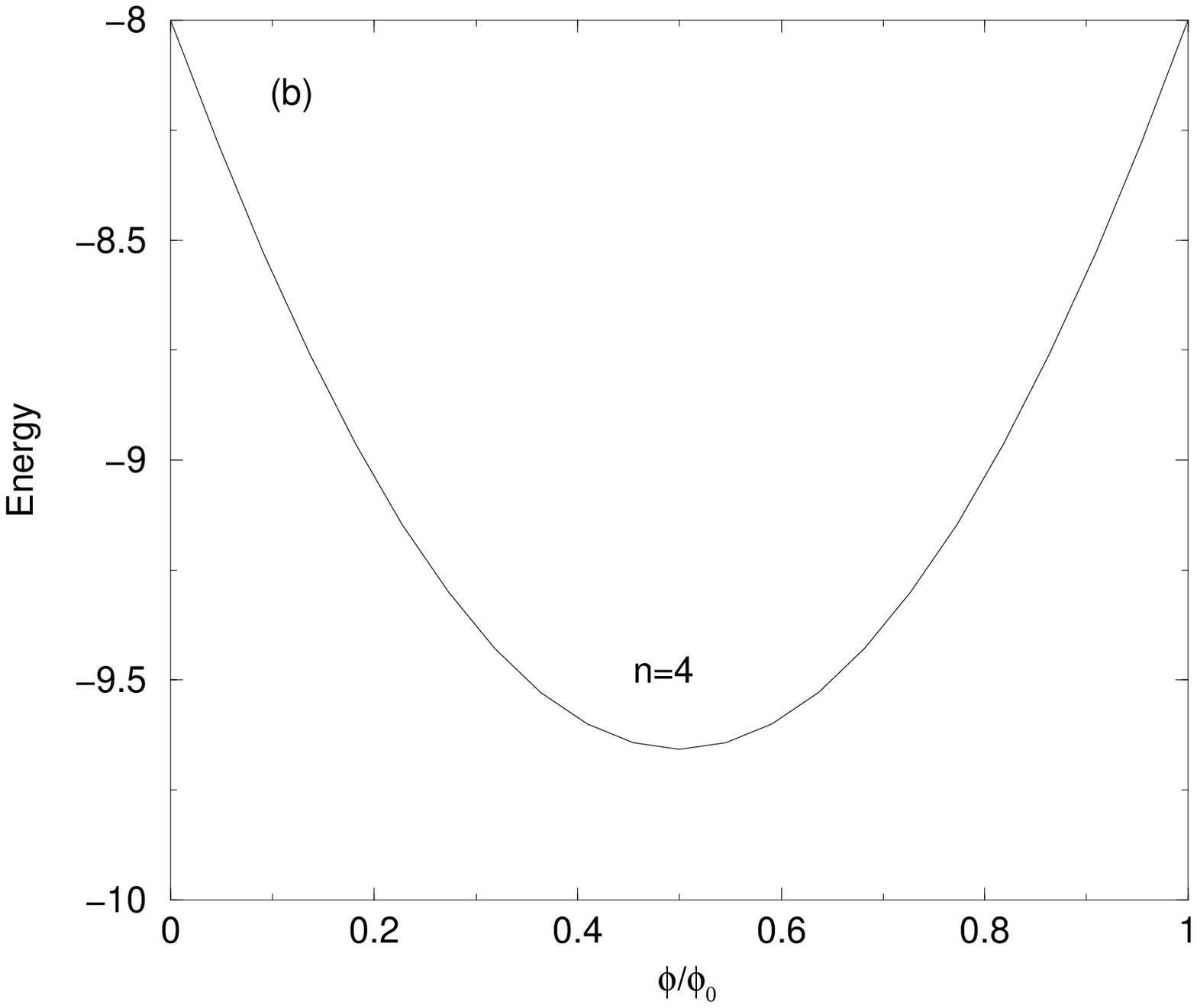,height=6cm,width=6cm}
\psfig{file=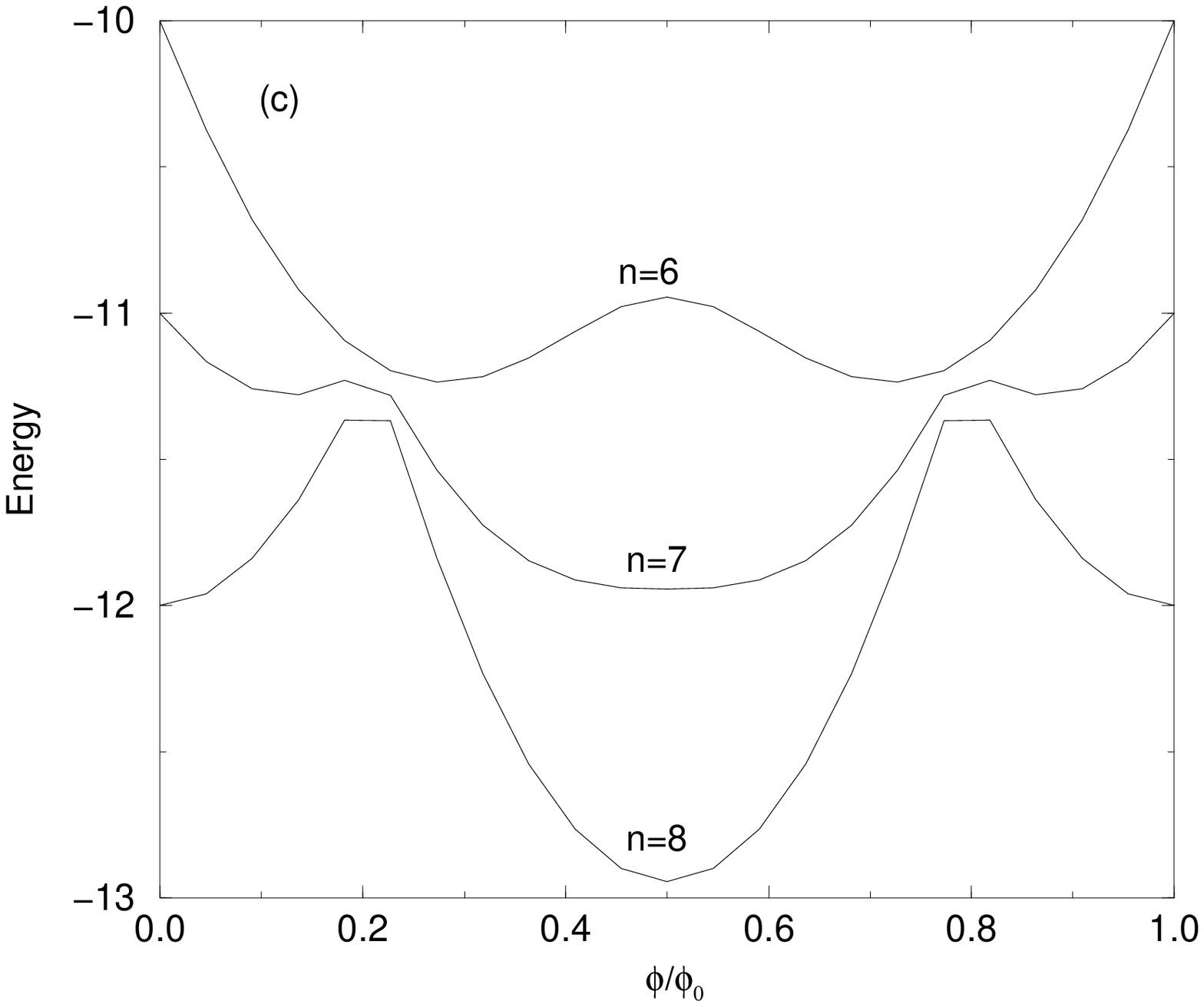,height=6cm,width=6cm}
\caption{Dependence of the ground state energy 
upon magnetic flux. All three interaction 
parameters are zero, i.e. $U=W=V=0$.}
\end{figure}
\newpage
\begin{figure}
\psfig{file=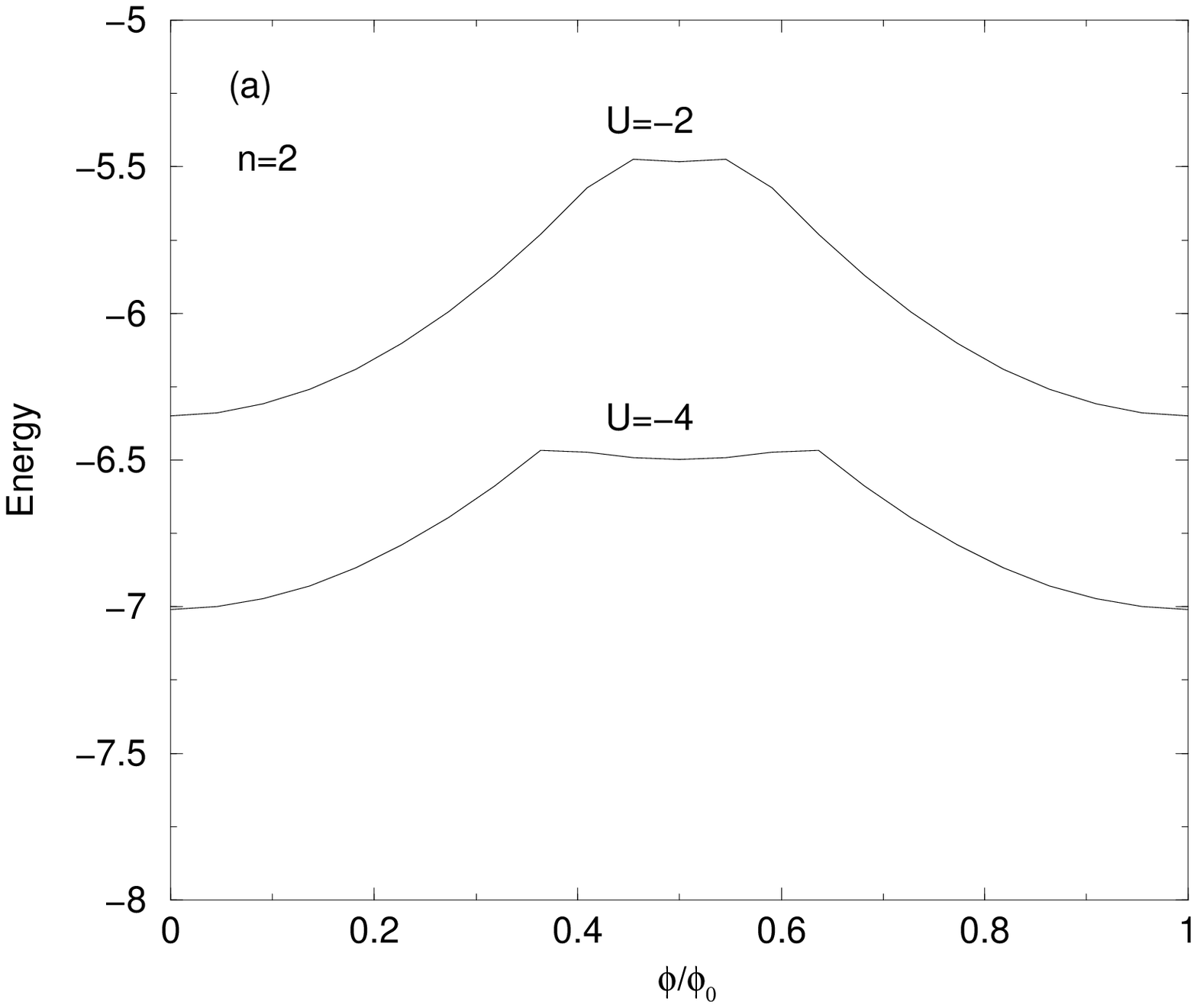,height=6cm,width=6cm}
\psfig{file=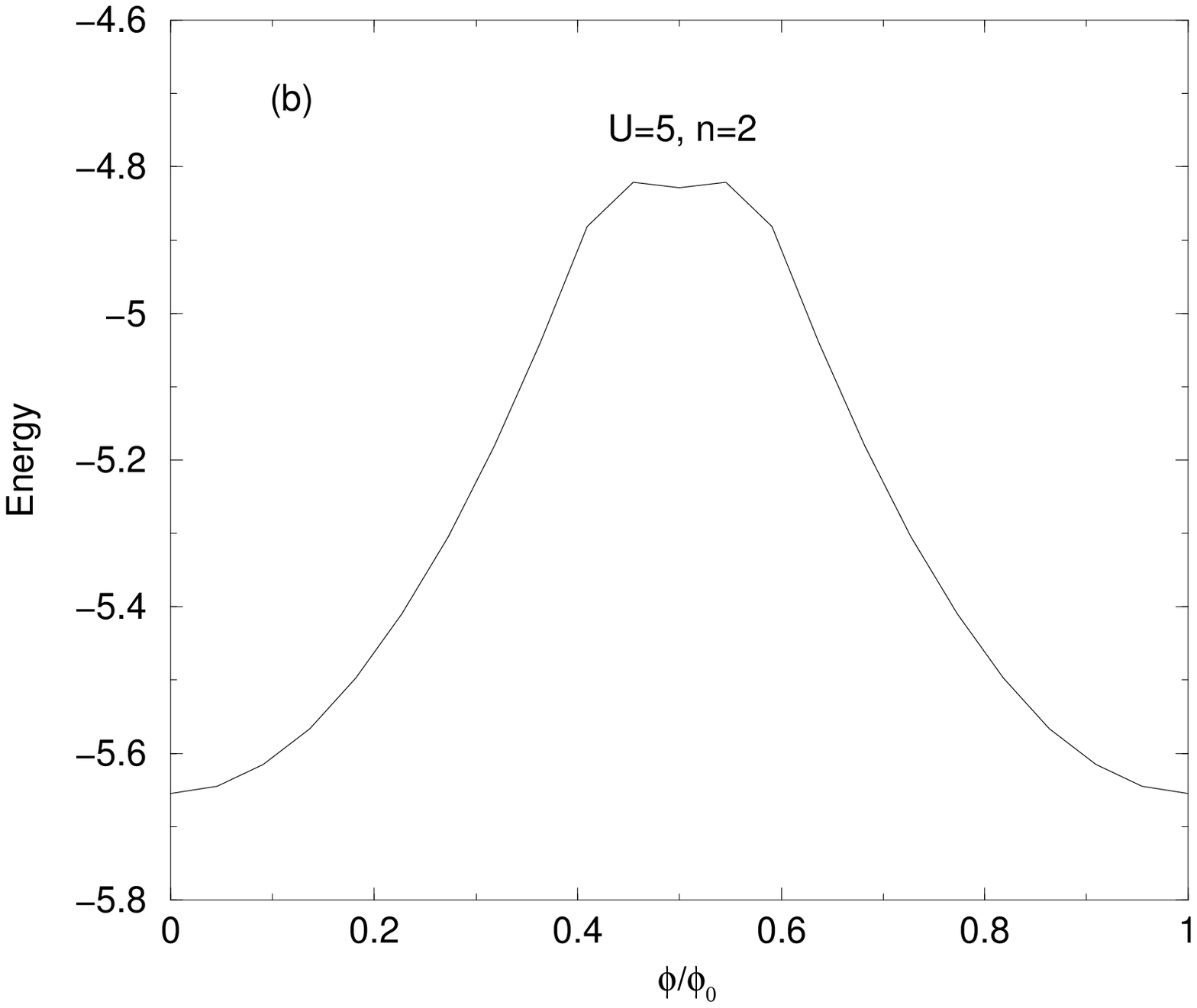,height=6cm,width=6cm}
\caption{Dependence of the ground state energy upon magnetic flux.
Contraction parameters are both zero, i.e. $W=V=0$, only the on-site
interaction parameter $U$ is nonzero.}
\end{figure}
\newpage
\begin{figure}
\psfig{file=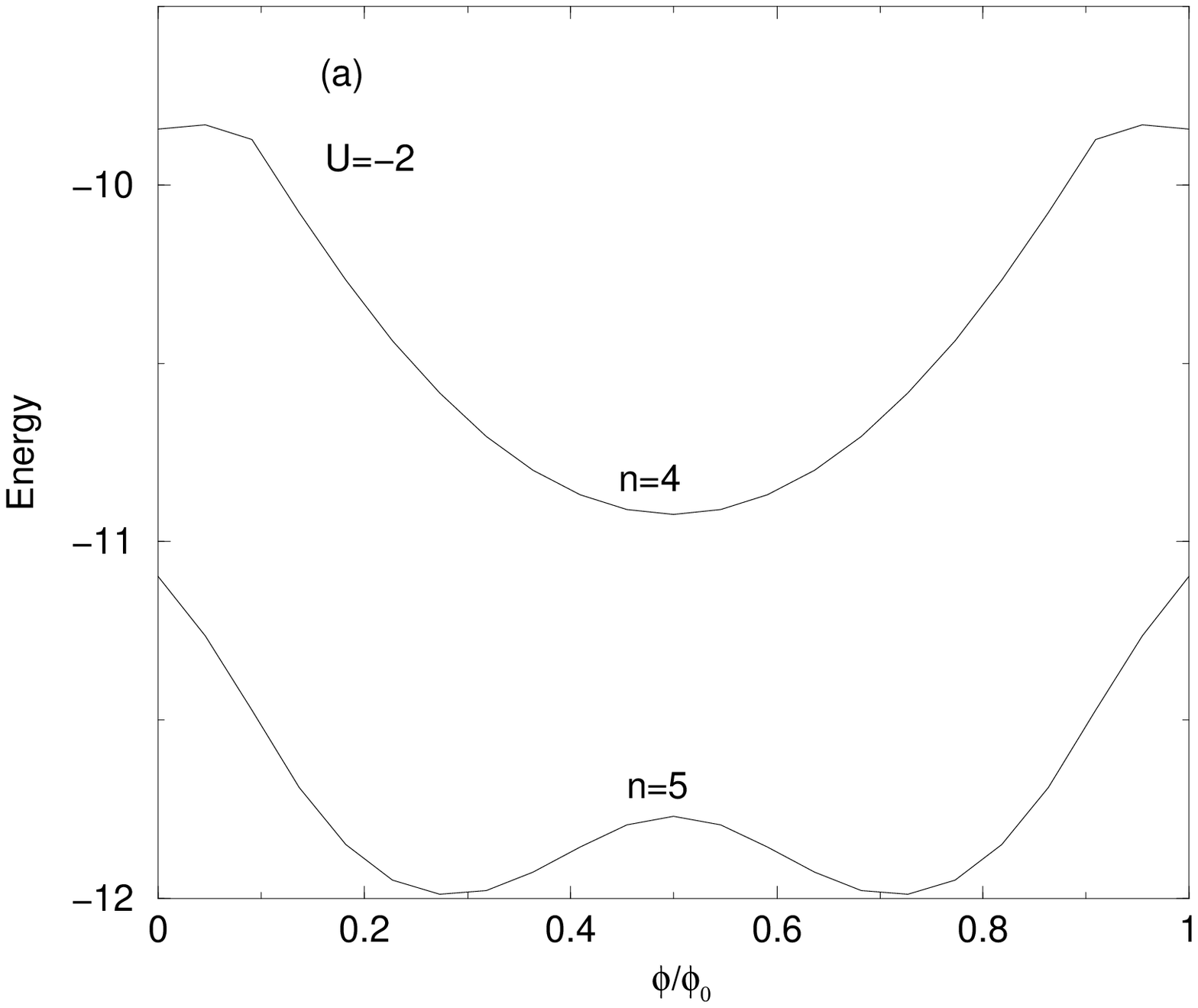,height=6cm,width=6cm}
\psfig{file=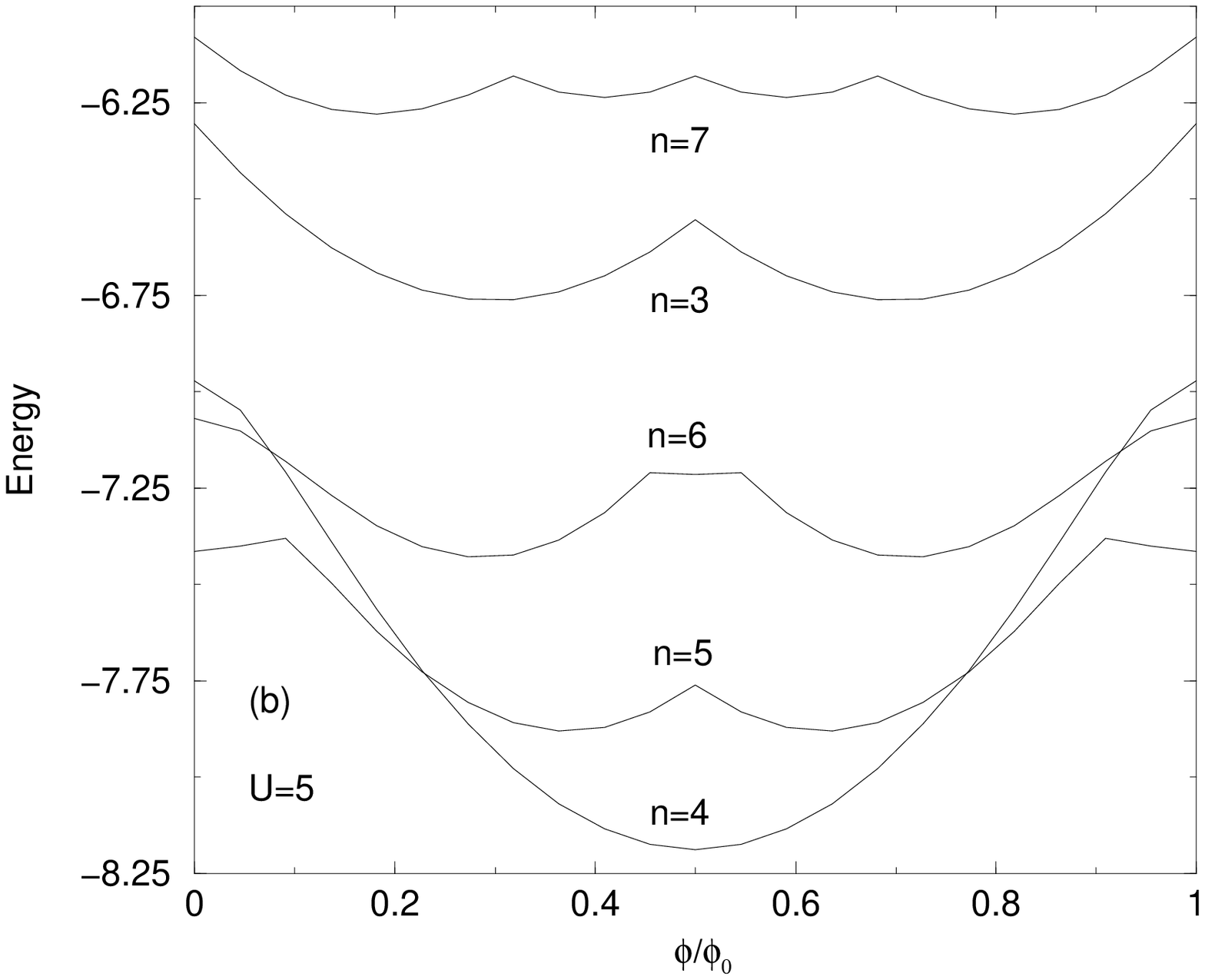,height=6cm,width=6cm}
\caption{Dependence of the ground state energy upon magnetic flux. 
Comparing (a) with Figure 9(b) clearly shows that the change in the 
parity of the number of particles for the case of negative $U$ values
introduces a sign change in the slope of $E(\Phi)$ at $\Phi=0$.
}
\end{figure}
\newpage
\begin{figure}
\psfig{file=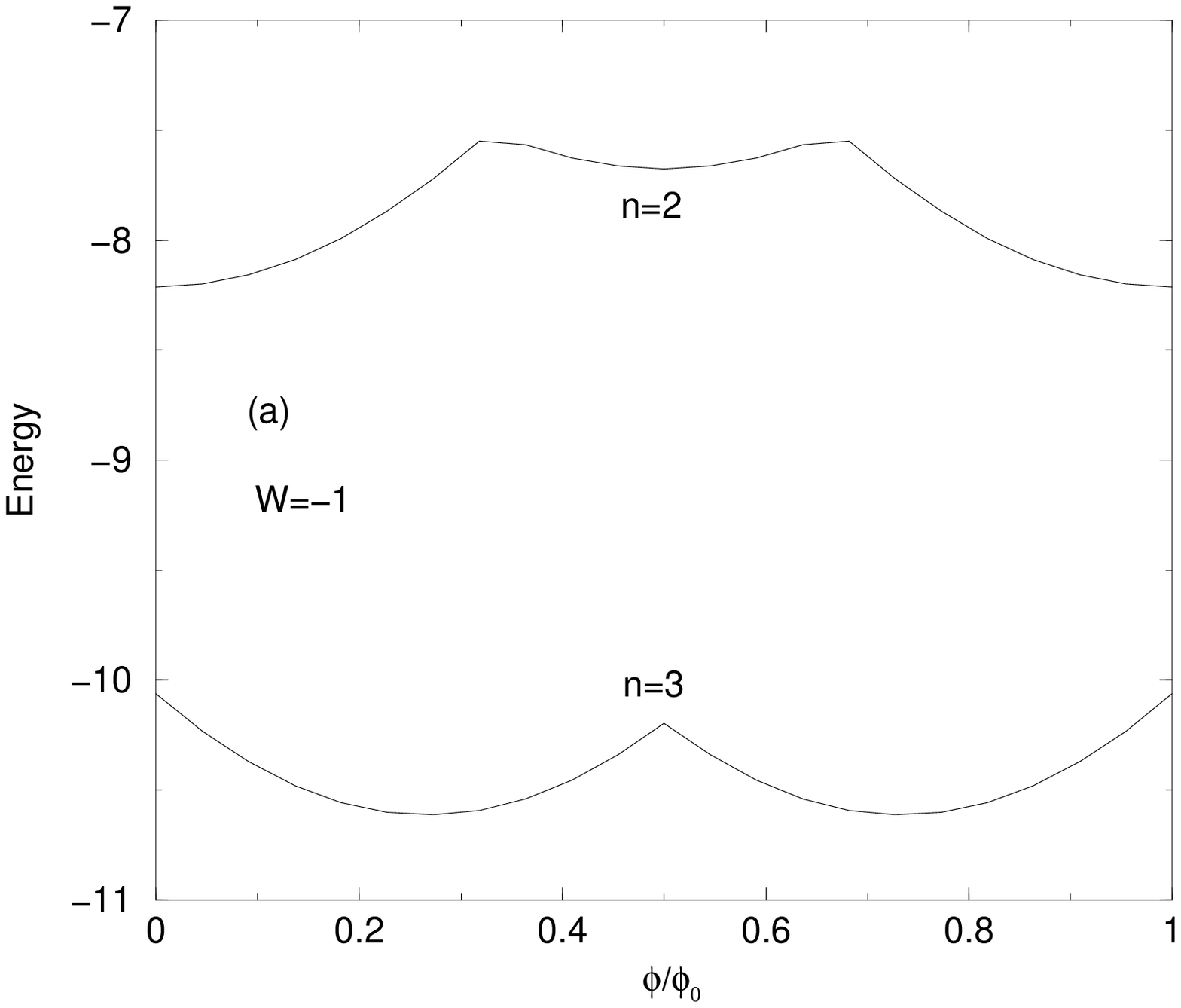,height=6cm,width=6cm}
\psfig{file=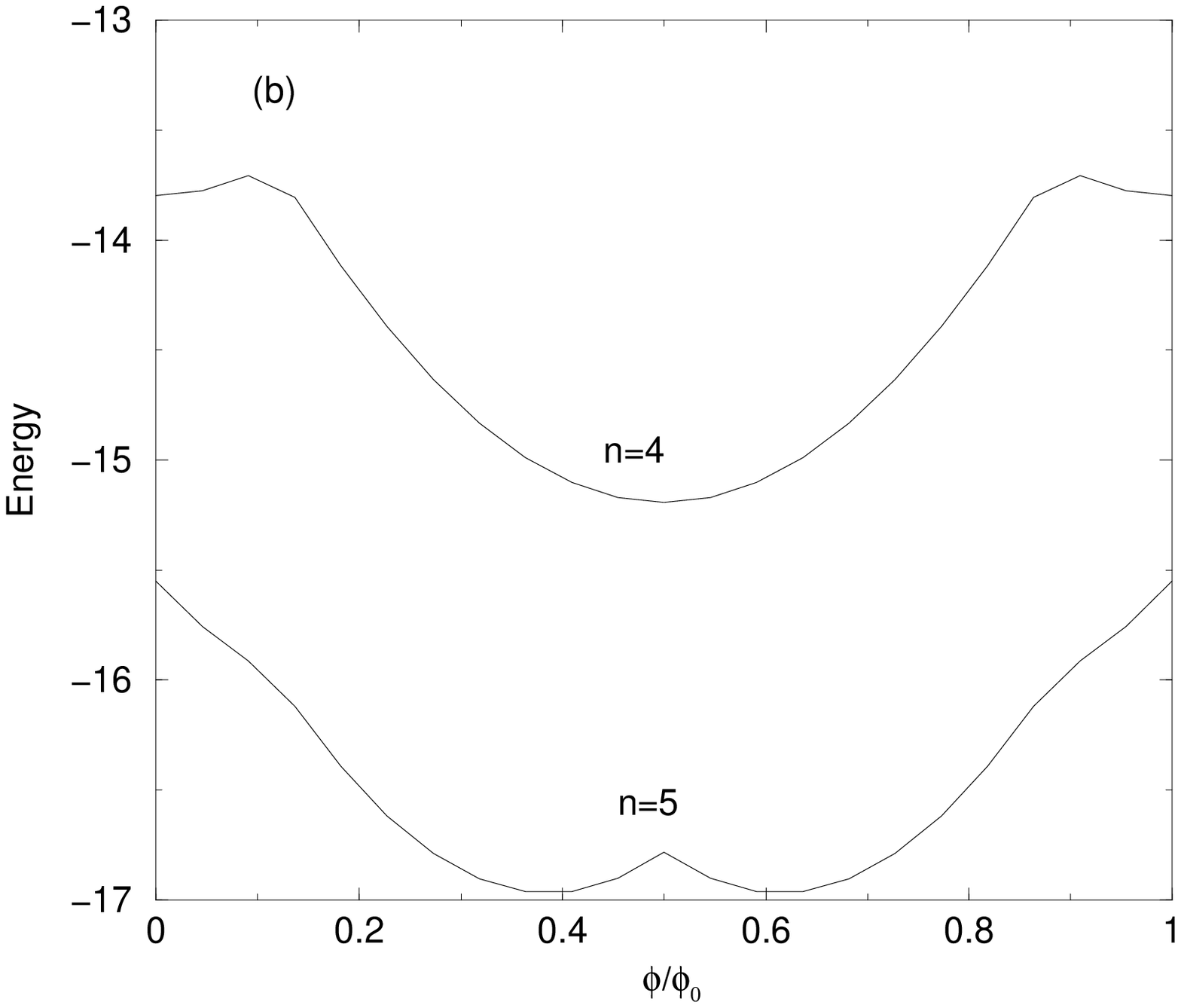,height=6cm,width=6cm}
\psfig{file=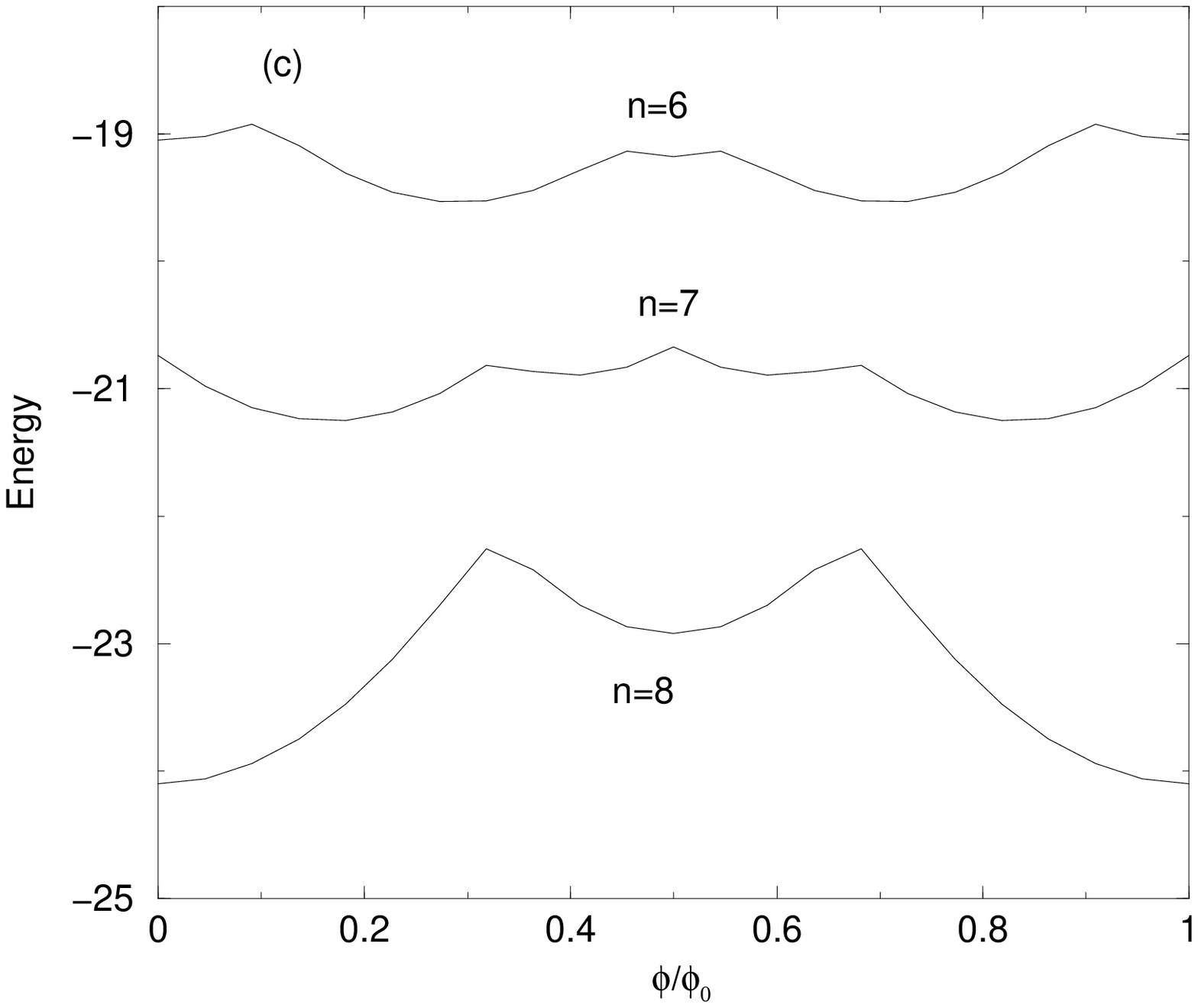,height=6cm,width=6cm}
\caption{Dependence of the ground state energy upon the magnetic flux. On-site
interaction parameter $U$ and one of the contraction parameters, $V$, 
are zero. All plots correspond to the non-zero interaction parameter $W=-1$. }
\end{figure}

\begin{table}
\begin{tabular}{|c|c|c|c|c|} \hline
 & \multicolumn{2}{c|} {$U=W=0$} & \multicolumn{2}{c|} {$U=V=0$} \\ \cline{2-5}
 & $V\rightarrow 0^+$ & $V\rightarrow 0^-$ &
$W\rightarrow 0^+$ & $W\rightarrow 0^-$ \\ \hline
below half-filling & & & & \\
($\mu<0$) &  $\Delta=0$ (no pairing) & $\Delta \neq 0$ (pairing)
& $\Delta=0$ (no pairing) & $\Delta \neq 0$ (pairing) \\ \hline
above half-filling & & & & \\
($\mu>0$) &  $\Delta \neq 0$ (pairing) & $\Delta=0$ (no pairing)
& $\Delta \neq 0$ (pairing) & $\Delta=0$ (no pairing) \\ \hline
\end{tabular}

\vspace{0.5cm}
\caption{Pairing effect for arbitrarily small values of $V$ and $W$,
computed by exact diagonalization of the Hamiltonian. The
results presented here are in complete agreement
with the
perturbative calculations.}
\end{table}

\end{document}